\documentclass[11pt]{article}
\pdfoutput=1
\usepackage{jheppub}

\usepackage[svgnames,table]{xcolor}
\usepackage{graphicx}
\usepackage{caption}
\usepackage{subcaption}
\usepackage{amsfonts,amsmath,amssymb,amsthm}
\usepackage{mathrsfs,bm,bbm,esint}
\usepackage[mathscr]{eucal}
\usepackage{physics}
\usepackage{dsfont}
\usepackage{slashed,soul}
\usepackage{rotating,pdflscape}
\usepackage{enumerate}
\usepackage{multirow,array}
\usepackage[numbers,sort&compress]{natbib}
\usepackage{tikz}
\usetikzlibrary{cd}
\usetikzlibrary{decorations.pathmorphing}
\usetikzlibrary{decorations.pathreplacing}
\usetikzlibrary{decorations.markings}
\tikzset{snake it/.style={decorate, decoration=snake}}
\tikzset{->-/.style={decoration={
  markings,
  mark=at position .5 with {\arrow{>}}},postaction={decorate}}}
\usepackage{subcaption}
\usepackage{float}
\usepackage{afterpage}
\renewcommand{\thefigure}{\arabic{figure}}

\title{A study of non-linear Langevin dynamics under non-Gaussian noise with quartic cumulant}
 
 \author[a,b]{Chandan Jana}

\affiliation[a]{Mandelstam Institute for Theoretical Physics, Witwatersrand University, Johannesburg, South Africa.}
\affiliation[b]{International Centre for Theoretical Sciences, Hesaraghatta, Bangalore, India.}

\emailAdd{channdann.jana@gmail.com }

\abstract{We consider a non-linear Langevin equation in presence of non-Gaussian noise originating from non-linear bath. We claim, the parameters in the Langevin equation are not physical. The physical parameters are obtained from a path-integral description of the system, where the Langevin parameters are related to the physical parameters by renormalisation flow equations. Then we compute both numerically and analytically the velocity two point function and show that it saturates to the bath temperature even in presence of non-linearity. We also find the velocity four point function numerically and show that it saturates to the analytically evaluated thermal velocity four point function when the non-linear FDR \cite{Chakrabarty:2019qcp} is satisfied. When the non-linear FDR, which is a manifestation of time reversibility of thermal bath, is violated then the system does not seem to thermalise. Rather, its velocity four point function settles to a steady state.}

\begin{document}
\maketitle
\raggedbottom

\section{Introduction}
\label{sec:intro}

A Brownian particle interacting with its thermal bath is successfully described by linear Langevin equation. The particle satisfies (linear) fluctuation-dissipation relation (FDR) which states that the variance of bath fluctuation is directly proportional to linear drag of the bath. The linear Langevin equation can be derived from a microscopic theory (Caldeira-Leggett model \cite{Caldeira1982PathIA}) of the particle interacting with a harmonic bath by a bilinear interaction. 
Adding a weak non-linear system-bath interaction \cite{Chakrabarty:2019qcp,Chakrabarty:2018dov} in the microscopic description is expected to modify the linear Langevin equation by certain non-linear terms.

The authors of \cite{Chakrabarty:2019qcp} consider a particle interacting with a thermal, harmonic bath at inverse temperature $\beta$ via a cubic interaction of the form: $\sum_{i,j} \lambda_{ij}\, q\, x_i x_j$. Here, $q$ and $x_i$ are system and bath degrees of freedom respectively, coupled via $\lambda_{ij}$.  They integrate out the bath perturbatively assuming small $\lambda_{ij}$ in Schwinger-Keldysh (SK) path-integral formalism \cite{Schwinger:1960qe,Keldysh:1964ud}\footnote{A textbook treatment of Schwinger-Keldysh prescription and its application to non-equilibrium systems can be found in \cite{Calzetta:1986ey,Kamenev}.} and find a quartic effective action for the particle. Along with the linear FDR, they find a \textbf{non-linear FDR} at the level of non-linear effective parameters. 

Identical FDRs are also derived by integrating out anharmonic, strongly coupled bath (in contrast to harmonic bath) for a particle \cite{Chakrabarty:2019aeu} and scalar field \cite{Jana:2020vyx}. More specifically, the baths in these models are conformal field theories (CFT) at non-zero temperature,\footnote{A conformal field theory (CFT) at zero temperature does not have any scale. But, a CFT at non-zero temperature has only one scale which is the temperature itself.}\footnote{A strongly interacting conformal bath is in general hard to integrate out. We circumvent this by using the duality between the anti-de Sitter (AdS) space and conformal field theory (CFT) \cite{Aharony:1999ti}. According to AdS/CFT duality the strongly interacting CFT bath is mapped to a weakly coupled AdS bath. So, we can integrate out the bath in AdS perturbatively using the gravitational SK prescription \cite{Glorioso:2018mmw,Chakrabarty:2019aeu,Jana:2020vyx}.} interacting linearly with the system.
So, we notice that the FDRs do not depend on specific details of the bath - whether a bath being harmonic or strongly interacting CFT. The FDRs are also not restricted to appear only at the level of quartic effective action; those are more generic. One can model CFT baths \cite{Jana:2020vyx} such that any specific higher degree interaction terms such as $q^n$ are generated in contrast to only quartic terms as found in \cite{Chakrabarty:2019qcp}. Surprisingly, the parameters corresponding to higher degree interaction terms follow identical non-linear FDR. Hence the non-linear FDR appears to be fairly generic; it only requires the bath to be thermal and time reversal invariant \cite{Berges:2000ur,Chakrabarty:2019aeu,Jana:2020vyx,Chakrabarty:2019qcp}. Therefore, it is worthwhile to test the implication of the FDRs in thermalisation. In particular, whether the velocity variance for a Brownian particle in a non-linear theory still saturates to the temperature (like in a linear theory by assuming the Boltzmann constant $k_B=1$ in suitable unit) or whether there is a systematic violation. Moreover, how are the higher point velocity correlations affected by the FDRs?

The non-linear effective theory and FDRs can have potential application in cosmology. Small non-Gaussian features (non-zero higher point correlation) in the cosmic microwave background (CMB) have been estimated (\cite{Maldacena:2002vr,ALLEN198766,Gangui:1993tt,Acquaviva:2002ud,PhysRevD.46.4232} and references therein) using inflationary cosmological models. The system field in these models is in a thermal bath of field(s) which live outside the cosmological horizon. The non-Gaussian features can be mapped to a non-linear stochastic problem \cite{10.1007/3-540-16452-9_6} and consequently the non-linear FDR could play an important role for a better interpretation of the CMB data. We leave it for future work. The FDRs may also appear in the effective theory of gravitating (compact, astrophysical) objects \cite{Goldberger:2004jt,Galley:2009px}. The effective theory can be thought of as an open system in a bath of radiation. Since the radiation gravitons interact non-linearly with the compact object, we expect non-linear FDR to emerge. Similar to the above applications, many other weak non-linear open systems are candidates to possess the FDRs.

In this work, we study the effect of the FDRs in correlation functions by considering a simple quartic model of a non-linear Brownian particle. The quartic, position-translation invariant SK effective action \cite{Chakrabarty:2019qcp,Chakrabarty:2019aeu} of the particle is given by
\begin{equation}\label{eqn:SK_effective_action}
\begin{split}
S_{SK} = -\int dt \left\{ q_d \ddot{q} + \gamma^r q_d \dot{q} - i \frac{ f^r}{2} q_d^2 - i \zeta_\eta^r \frac{q_d^4}{4!} - \zeta_{\gamma}^r q_d^3\dot{q} \right\} \,,
\end{split}
\end{equation}
where $q,\,q_d$ are the Schwinger-Keldysh classical and fluctuation degrees of freedom respectively. The translation invariance in position allows only the time derivative of $q$ terms in the action. $\gamma^r,f^r,\zeta_\eta^r,\zeta_\gamma^r$ are the effective parameters emerged from the system-bath interaction. The significance of $r$-label on the parameters will become clear shortly.

The SK path integral with action \eqref{eqn:SK_effective_action} is dual to a classical Langevin equation following Martin-Siggia-Rose (MSR) \cite{Martin:1973zz,DeDominicis:1977fw,article} prescription (discussed in \S\ref{sec:renorm_parameter}). The dual Langevin equation is given by \cite{Chakrabarty:2019qcp,Chakrabarty:2019aeu}
\begin{equation}\label{eqn:nonLin_LangevinEqn}
\mathcal{E}(\dot{q},\eta) \equiv \ddot{q} + \left(\gamma+\zeta_{\gamma} \eta^2 \right) \dot{q} - f\, \eta = 0\,.
\end{equation}
Here $q$ is the particle's position. The parameters $\gamma$ and $f$ are damping coefficient and strength of the thermal noise respectively. The damping coefficient is corrected by $\zeta_\gamma \eta^2$ - a thermal jitter in $\gamma$ with strength $\zeta_\gamma$. The duality also ensures that the noise is drawn from a non-Gaussian distribution as the following.
\begin{equation}\label{eqn:nonG_noise}
P(\eta) \sim e^{- \delta t \left\{\frac{ f }{2} \eta^2 + \frac{\zeta_\eta}{4!} \eta^4 \right\}} \,,
\end{equation}
$\zeta_\eta$ being the strength of quartic noise. The above distribution is solely determined by the system-bath interaction; adding potential terms to \eqref{eqn:nonLin_LangevinEqn} does not affect the distribution. This means, e.g., if we study the Langevin equation \eqref{eqn:nonLin_LangevinEqn} numerically with a Gaussian noise, the particle should not thermalise (see figure \ref{fig:vel_var}). So, the non-Gaussian noise plays a role in particle's correlations functions.

Note that the parameters in \eqref{eqn:nonLin_LangevinEqn} and \eqref{eqn:nonG_noise} are not same as those appearing in \eqref{eqn:SK_effective_action}. We show in \S\ref{sec:renorm_parameter} that the Langevin parameters are related to the respective $r$-labelled SK parameters by a renormalisation flow, where the SK parameters are the renormalised/physical parameters and the Langevin parameters are the bare parameters. The renormalised parameters in the SK description follow certain relations among themselves as the following:
\begin{equation}\label{eqn:FDRs}
\gamma^r = \frac{\beta}{2} f^r  \qquad \text{and} \qquad \zeta_\gamma^r = - \frac{\beta}{12} \zeta_\eta^r \,.
\end{equation}
The first relation can be identified as the linear FDR between the renormalised damping coefficient $\gamma^r$ and the renormalised strength of the noise $f^r$. $\beta$ is the inverse temperature of the bath. The second relation is a non-linear generalisation of the linear FDR, thus can be called a non-linear FDR \cite{Wang:1998wg,Chakrabarty:2019qcp,Chakrabarty:2019aeu,Jana:2020vyx}. It states that the renormalised coefficient $\zeta_\gamma^r$ is linearly related to the renormalised non-Gaussian parameter $\zeta_\eta^r$. The linear FDR requires the bath to be thermal whereas the non-linear FDR also needs the thermal bath to be time reversal invariant \cite{Chakrabarty:2019qcp,Chakrabarty:2018dov}. The explicit numerical value of $\zeta_\eta^r$ as derived in \cite{Chakrabarty:2019aeu,Jana:2020vyx} turns out to be negative which makes the noise distribution ill defined. So, we regulate the noise distribution (\S\ref{subsec:preperation_noise}) by a cut-off.

In the first part of this work, we use the SK path integral to compute the velocity two and four point functions analytically in \S\ref{sec:corr_analytic}. Since \eqref{eqn:SK_effective_action} is an interacting SK action, we choose $\zeta^r_\eta,\zeta^r_\gamma$ to be sufficiently small to implement perturbation theory. However, due to lack of available analytic tools, we evaluate only thermal correlations analytically and match against numeric.

On the other side, we consider the Langevin description for a numerical study of velocity correlations in \S\ref{sec:numerical_analysis}. In the Langevin description, first we prepare the non-Gaussian noise \eqref{eqn:nonG_noise} using a numerical tool called the \textbf{rejection sampling} \cite{vonNeumann1951}. According to this tool one can prepare an arbitrary noise distribution from a known distribution. Given the required non-Gaussian noise distribution, we solve \eqref{eqn:nonLin_LangevinEqn} numerically and find that the velocity variance for a non-linear Brownian particle saturates to $\frac{1}{\beta}$ - the bath temperature. We expect that the connected velocity four point function (equal time) to be non-zero due to presence of non-linearity. The numerically evaluated four point function is indeed non-zero and it saturates after some characteristic time. We show that it thermalises (by checking against analytic result) only if the non-linear FDR is valid. Otherwise, the particle reaches a steady state\cite{2002cond.mat..2501M,Harris_2007,Parrondo_2009,Andrieux_2009,PhysRevLett.101.090602}. This means, thermal nature of the bath is not sufficient to guarantee thermalisation of a non-linear Brownian particle; it also requires time-reversal invariance\cite{Berges:2000ur}.


\section{Renormalisation of parameters in the discretised problem}
\label{sec:renorm_parameter}

The relations between the SK and the Langevin parameters will be central in matching analytic correlations against the numeric ones. So, we begin by deriving those relations following the Martin-Siggia-Rose (MSR) prescription. The MSR prescription establishes a duality between SK path-integral and stochastic description of a system. This means, each Langevin equation has a SK path integral interpretation. The parameters appearing in the SK description are physical, thus we call those as \textbf{renormalised} parameters. The parameters appearing in the Langevin equation are related to renormalised ones by renormalisation flow equations. We call those as the \textbf{bare} parameters. In this section, we find relations among the bare and renormalised parameters. 

The duality between the Langevin and path integral description is established by the following steps
\begin{itemize}
\item First we write an identity for the noise path-integral: $1 =\mathcal{Z} =  \int \mathcal{D}\eta \, P[\eta]$ where $\eta$ is the noise and $P[\eta]$ is the distribution functional for $\eta$.
\item Then we introduce the Langevin equation by a Dirac-Delta function as $$\mathcal{Z} =  \int \mathcal{D}\eta \mathcal{D}q \,P[\eta]\,  \text{det}\left[\frac{\delta \mathcal{E}}{\delta q}\right] \, \delta(\mathcal{E}(q,\eta)),$$ where $\mathcal{E}(q,\eta)=0$ is the Langevin equation.
\item We write the exponential form of the Dirac-Delta function by introducing an auxiliary variable $q_d$ as the following.
$$ \delta(\mathcal{E}) = \int \mathcal{D} q_d \, e^{-i\int dt \, \mathcal{E}q_d} .$$
\item The noise $\eta$ is then integrated out preceded by a shift: $\eta \rightarrow \eta+iq_d$.
\item Finally we identify $q$ and $q_d$ to the SK classical and fluctuation degrees of freedom respectively. The final path-integral is identified to the SK path-integral which is unity in absence of any sources.
\end{itemize}
Following the above steps of MSR rules, the path integral of the non-linear Langevin particle \eqref{eqn:nonLin_LangevinEqn} with noise distribution \eqref{eqn:nonG_noise} is given by
\begin{equation}\label{eqn:partition_fn0}
\begin{split}
1 = \mathcal{Z} =&  \int \mathcal{D}q \, \mathcal{D}\eta \, \det\left[ \frac{\delta \mathcal{E}}{\delta q} \right] \, P[\eta] \, \delta\left( \mathcal{E}(t) \right) \\
=& \int \mathcal{D}q \, \mathcal{D}\eta \, \mathcal{D} q_d\, \det\left[ \frac{\delta \mathcal{E}}{\delta q} \right] \, e^{-\int dt \left\{\left[ \frac{f}{2} \eta^2 + \frac{\zeta_\eta}{4!} \eta^4 \right] + i q_d \mathcal{E}(t) \right\}} \,.
\end{split}
\end{equation}
Here $q_d$ is the difference/fluctuation d.o.f. in SK prescription. We follow the Stratonovich convention\cite{cite-key1,Stratonovich1992NonlinearNT} and write the determinant in \eqref{eqn:partition_fn0} as a path integral over two Grassmannian variables ($\bar{c},c$) as the following.\footnote{One could have also used the Ito convention \cite{cite-key,Kamenev} and set the determinant to unity. In doing so we will end up making the $\zeta_\gamma$ term non-local \cite{Chaudhuri:2020sc} by one time step $\delta t$. Correspondingly, the path integral description becomes non-local which is cumbersome to handle.} 
\begin{equation}\label{eqn:partition_det}
\begin{split}
\det\left[ \frac{\delta \mathcal{E}}{\delta q} \right] 
=& \int  \mathcal{D} \bar{c} \,\mathcal{D}c\,\, e^{-i \int dt\, \bar{c} \left[ \frac{d}{dt} + \gamma +\zeta_\gamma \eta^2 \right] c }  \,.
\end{split}
\end{equation}

We give a shift to the noise by $\eta\rightarrow \eta + i q_d $. The $q_d$ can be thought of as a dual variable corresponding to the noise and is identified to the fluctuation d.o.f. in SK description. The shifting does not essentially change the path integral:
\begin{equation}\label{eqn:partition_fn_shifted}
\begin{split}
\mathcal{Z} =&  \int \mathcal{D}q \, \mathcal{D}\eta \, \mathcal{D} q_d \, \mathcal{D} \bar{c} \mathcal{D}c\, \, e^{-\int dt \left\{\left[ \frac{f}{2} (\eta+iq_d)^2 + \frac{\zeta_\eta}{4!} (\eta+iq_d)^4 \right] + i q_d \mathcal{E}(t) + i\, \bar{c} \left[ \frac{d}{dt} + \gamma +\zeta_\gamma (\eta+iq_d)^2 \right] c\right\}} \,.
\end{split}
\end{equation}
%
Then we integrate out the noise by computing noise-noise correlation functions. Since there is non-Gaussianity in the noise distribution, we can only integrate out the noise perturbatively. The path integral in terms of noise correlation functions takes the following form in perturbation theory.
\begin{equation}\label{eqn:Ztailor_expnd}
\begin{split}
\mathcal{Z}
=&
\int \mathcal{D}q \mathcal{D}q_d \mathcal{D}\bar{c} \mathcal{D}c \, e^{-\int dt \left[ \frac{f}{2} q_d^2 + \frac{\zeta_\eta}{4!} q_d^4 +iq_d\left( \ddot{q} +(\gamma-\zeta_\gamma q_d^2)\dot{q} \right) + i\,\bar{c} \left( \frac{d}{dt} + \gamma - \zeta_\gamma q_d^2 \right)c  \right]} \\
&
 \Bigg\{ 1 + \int dt \left[ \frac{\zeta_\eta}{2!2!}\langle \eta^2\rangle q_d^2 - i \zeta_{\gamma}\langle \, \eta^2\rangle q_d \dot{q} -i \zeta_\gamma \langle \eta^2 \rangle \bar{c}c \right] + \frac{i\zeta_\gamma \zeta_\eta}{4!} \int dt\, dt' \, \langle \eta^4(t') \eta^2(t)\rangle q_d(t)\dot{q}(t) \\
& -
 \frac{\zeta_\eta^2}{4!} \int dt \,dt' \left[ \frac{1}{2^2} \langle \eta^2(t) \eta^4(t')\rangle \, q_d^2(t)  + \frac{1}{3} \langle \eta^3(t) \eta^3(t')\rangle \, q_d(t)q_d(t') \right] \\
& + 
\frac{\zeta_\eta^2}{4!} \int dt \,dt' \left[ \frac{2}{3} \langle \eta(t) \eta^3(t')\rangle \, q_d^3(t)q_d(t')  + \frac{3}{4} \langle \eta^2(t) \eta^2(t')\rangle \, q_d^2(t)q_d^2(t') \right] \\
& - i\zeta_\eta \zeta_\gamma  \int dt \,dt' \left[ \frac{1}{3} \langle \eta(t) \eta^3(t')\rangle \, q_d^2(t)\dot{q}(t) q_d(t')  + \frac{1}{4} \langle \eta^2(t) \eta^2(t') \rangle \, q_d^2(t)q_d(t')\dot{q}(t') \right] \\
& -
i \zeta_\gamma \zeta_\eta \int dt\, dt' \left[ \frac{1}{4} \langle \eta^2(t) \eta^2(t') \rangle \, \bar{c}(t)c(t) q_d^2(t') + \frac{1}{3} \langle \eta(t) \eta^3(t') \rangle \, \bar{c}(t)c(t) q_d(t) q_d(t') \right] +\cdots
 \Bigg\} \,.
\end{split}
\end{equation}
%
Note that, in the Taylor expanded form, we have kept terms till quartic order(in $q,q_d,\bar{c},c$). The higher order terms are suppressed due to small non-linear couplings. The noise correlation functions are UV divergent; thus we put a UV cut-off $\delta t$ in time which also acts as the smallest time step in the discretised problem discussed through out this paper.

We evaluate various noise correlation functions of \eqref{eqn:Ztailor_expnd} in appendix \ref{app:noise_corr}. With the noise correlation function, we exponentiate the RHS of \eqref{eqn:Ztailor_expnd} back and find the following effective path integral.
\begin{equation}\label{eqn:SK_effective_path_integral}
\begin{split}
\mathcal{Z} =
\int \mathcal{D}q\, \mathcal{D}q_d\, \mathcal{D}\bar{c} \, \mathcal{D}c 
\times \exp\Bigg[-\int dt \Bigg\{ i q_d \ddot{q} + i\gamma^r q_d \dot{q} + \frac{ f^r}{2} q_d^2 + i \bar{c}\left( \frac{d}{dt} + \gamma - \zeta_\gamma q_d^2 \right)c \\+ \zeta_\eta^r \frac{q_d^4}{4!} - i\zeta_{\gamma}^r q_d^3\dot{q}  + \cdots \Bigg\} \Bigg] \,.
\end{split}
\end{equation}
The $r$ labelled parameters are renormalised parameters (evaluated in appendix \ref{app:noise_corr}) and are given by 
\begin{subequations}\label{eqn:reln_bare_renorm_parameters}
\begin{align}
f^r &= f - \frac{\zeta_\eta}{2  f \delta t} + \frac{2\zeta_\eta^2 }{3 f^3 (\delta t)^2} + \cdots \,, \\
\gamma^r &= \gamma + \frac{\zeta_\gamma}{ f \delta t} - \frac{\zeta_\gamma \zeta_\eta}{2 f^3(\delta t)^2} + \cdots \,,\\
\zeta_\eta^r &= \zeta_\eta - \frac{7}{2}\frac{\zeta_\eta^2}{ f^2\delta t } + \frac{149}{12} \frac{\zeta_\eta^3}{ f^4 (\delta t)^2} +\cdots \,, \\
\zeta_\gamma^r &= \zeta_\gamma - \frac{3\zeta_\eta\zeta_\gamma}{2 f^2\delta t} + \frac{10}{3} \frac{\zeta_\eta^2\zeta_\gamma}{ f^4 (\delta t)^2} + \cdots \,.
\end{align}
\end{subequations}
Here $\delta t$ is the UV cut-off and the smallest time step and the perturbation parameter is $\sim\frac{\zeta_\eta}{f^2\delta t}$.
Note that the front factors, in RHS of the above relations, are larger down the series. So, we choose the bare parameters such that perturbation theory is valid. All of the renormalised parameters are not independent. The linear renormalised parameters $\{f^r,\gamma^r\}$, as discussed in the introduction, are related by the linear FDR:
\begin{equation}\label{eqn:LinFDR}
\gamma^r = \frac{\beta}{2} f^r  \,.
\end{equation}
If the bath also possesses a time reversal symmetry, the non-linear renormalised parameters are related by the non-linear FDR:
\begin{equation}\label{eqn:nonLinFDR}
\zeta_\gamma^r = - \frac{\beta}{12} \zeta_\eta^r \,.
\end{equation}

The parameters corresponding to the Grassmann variables $\bar{c},c$ also obtain correction from noise correlation function, but those can be neglected due to the following reason. Since, $\bar{c}$ and $c$ do not constitute any physical observable due to their Grassmannian nature, they do not have any direct contribution in any correlation function; they only have sub-leading loop contribution. So, the leading parameters corresponding to Grassmann vertices in \eqref{eqn:SK_effective_path_integral} make sub-leading contribution to correlation function. Hence noise loop correction to those parameters will be sub-sub-leading and thus are ignored (see appendix \ref{app:SK_prop}).

One might be concerned about LHS of \eqref{eqn:reln_bare_renorm_parameters} being called as renormalised parameters, since they may receive correction due to quartic terms in \eqref{eqn:SK_effective_path_integral} and be further renormalised. We will argue in the next section that the renormalised parameters are not further corrected by $\zeta_\gamma^r$ or $\zeta_\eta^r$ interactions. Although we get $\bar{c}$-$c$ loop corrections but those are sub-leading as discussed above. Therefore we can safely call LHS of \eqref{eqn:reln_bare_renorm_parameters} as the renormalised parameters.

The bare parameters in \eqref{eqn:reln_bare_renorm_parameters}, according to the notion of renormalisation, should flow with $\delta t$ whereas the renormalised parameters are fixed by the system itself. This means that the bare parameters can take different values along the flow depending on $\delta t$. One can find the flow equations by writing the bare parameters in terms of the renormalised ones, which can be found easily by making suitable approximation.

\section{Thermal velocity correlations in Schwinger-Keldysh formalism}
\label{sec:corr_analytic}

Correlation function for a generic initial condition is hard to compute analytically. In this section we rather compute thermal correlations such as velocity variance and equal time velocity four point correlation analytically using the SK path integral.

\subsection{Schwinger-Keldysh propagators and velocity variance}
\label{subsec:skProp}

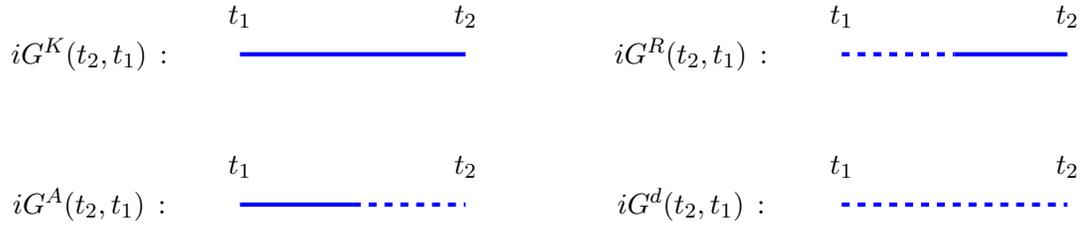
\begin{figure}
\centering
\begin{tikzpicture}
\begin{scope}[shift={(0,0)}]
\draw[blue, ultra thick] (2,0) -- (5,0);
\node at (0,0) {$iG^K(t_2,t_1)\, : $};
\node at (2,0.5) {$t_1$};
\node at (5,0.5) {$t_2$};

\draw[blue,ultra thick]  (2,-2) -- (3.5,-2);
\draw[blue,ultra thick,dashed]  (3.5,-2) -- (5,-2);
\node at (0,-2) {$iG^A(t_2,t_1)\,:$};
\node at (2,-1.5) {$t_1$};
\node at (5,-1.5) {$t_2$};
\end{scope}

\begin{scope}[shift={(8,0)}]
\draw[blue,ultra thick,dashed]  (2,0) -- (3.5,0);
\draw[blue,ultra thick]  (3.5,0) -- (5,0);
\node at (0,0) {$iG^R(t_2,t_1)\, : $};
\node at (2,0.5) {$t_1$};
\node at (5,0.5) {$t_2$};

\draw[blue, ultra thick,dashed] (2,-2) -- (5,-2);
\node at (0,-2) {$iG^{d}(t_2,t_1)\,:$};
\node at (2,-1.5) {$t_1$};
\node at (5,-1.5) {$t_2$};

\end{scope}
\end{tikzpicture}
\caption{Diagrammatic representation of Schwinger-Keldysh propagators: The explicit expression for the propagators are the following. $iG^K(t_2,t_1) = \langle \dot{q}(t_2) \dot{q}(t_1) \rangle \,,
iG^A(t_2,t_1) = \langle q_d(t_2) \dot{q}(t_1) \rangle \,,
iG^R(t_2,t_1) = \langle \dot{q}(t_2) q_d(t_1) \rangle $ and $
iG^{d}(t_2,t_1) = \langle q_d(t_2) q_d(t_1) \rangle$. The solid and dashed lines correspond to $\dot{q}$ and $q_d$ respectively.}
\label{fig:SK_propagators}
\end{figure}

We evaluate the thermal SK propagators (explicitly in appendix \ref{app:SK_prop}) to compute the velocity correlation functions. The SK propagators are given by
\begin{eqnarray}
\label{eqn:SK_propagatorK}
iG^K(t_2,t_1) \equiv && \langle \dot{q}(t_2) \dot{q}(t_1) \rangle = \frac{f^r}{2\gamma^r} e^{-\gamma^r|t_2-t_1|}  \,, \\
\label{eqn:SK_propagatorA}
iG^A(t_2,t_1) \equiv && \langle q_d(t_2) \dot{q}(t_1) \rangle = -i\, \Theta(t_1-t_2) e^{-\gamma^r (t_1-t_2)} \,, \\
\label{eqn:SK_propagatorR}
iG^R(t_2,t_1) \equiv && \langle \dot{q}(t_2) q_d(t_1) \rangle =  -i \, \Theta(t_2-t_1) e^{-\gamma^r (t_2-t_1)} \,, \\
\label{eqn:SK_propagatorD}
iG^{d}(t_2,t_1) \equiv && \langle q_d(t_2) q_d(t_1) \rangle =  0 \,.
\end{eqnarray}
Here, $\dot{q}$ is the particle's velocity. The above non-zero propagators are the Keldysh, advanced and the retarded propagators respectively. We depict the propagators in figure \ref{fig:SK_propagators}.

Note that the propagator in \eqref{eqn:SK_propagatorD} is zero. In absence of this propagator and with the available interaction terms in \eqref{eqn:SK_effective_path_integral}, there will not be any loop correction consisting of $q_d,\dot{q}$ to velocity variance. Although there will be Grassmann loop corrections, those are negligibly small, thus can be ignored (see appendix \ref{app:SK_prop}). Therefore, in absence of any non-linear correction, the Keldysh propagator \eqref{eqn:SK_propagatorK} is the velocity two point function and at equal time is the velocity variance. Thus the velocity variance, by using the linear FDR, is given by
\begin{equation}
\langle \dot{q}^2(t) \rangle = \frac{f^r}{2\gamma^r} = \frac{1}{\beta} \,.
\end{equation}
So, we find that the velocity variance is equal to the temperature of the bath, like in linear Langevin dynamics. Weak non-linearity does not bring in correction to velocity variance in thermal equilibrium.

\subsection{Microscopic time reversibility of bath and non-linear FDR}
\label{subsec:equal_v4}

The non-linear FDR, as discussed in the introduction, seems to be generic from the discussions of \cite{Chakrabarty:2018dov,Chakrabarty:2019qcp} and \cite{Chakrabarty:2019aeu}. In \cite{Chakrabarty:2018dov,Chakrabarty:2019qcp}, non-linear FDR was derived assuming weak non-linear system-bath interaction. On the other hand, \cite{Chakrabarty:2019aeu} found an identical FDR with linear system-bath interaction where the bath being non-linear and strongly coupled. Thus non-linear FDR does not seem to depend on the details of bath, rather it depends on a generic principle: microscopic time reversal invariance of the bath.

Microscopic time reversibility of bath, in general, imposes certain constraints on the system's effective theory. These relations were first discovered by Onsager \cite{Onsager:1931jfa,Onsager:1931kxm} and extended by Casimir \cite{RevModPhys.17.343} in the context of a system with multiple degrees of freedom. According to this observation, if $a)$ the bath Hamiltonian $H_B$ commutes with the time reversal operator $\textbf{T}$ and $b)$ the bath operator $\mathcal{O}$, through which the bath interacts with the system, is even under time reversal:
\begin{equation}\label{eqn:time_reversal_HO}
\begin{split}
\left[ \textbf{T},H_B \right] = 0 \,, \qquad  \textbf{T}^{-1}\, \mathcal{O}(t) \,\textbf{T} =  \mathcal{O}(-t)\,,
\end{split}
\end{equation}
then certain parameters in the system's effective theory are related. 

Following \cite{Chakrabarty:2018dov,Chakrabarty:2019qcp}, we motivate the above assertion in the context of non-linear FDR by considering the following bath four point function:
\begin{equation}
\langle \mathcal{O}(t_1) \mathcal{O}(t_2) \mathcal{O}(t_3) \mathcal{O}(t_4) \rangle \,.
\end{equation}
This kind of four point functions would appear in the influence phase of the effective theory when we integrate out the bath perturbatively. Due to time reversal invariance we have
\begin{equation}
\langle  \mathcal{O}(t_1) \mathcal{O}(t_2) \mathcal{O}(t_3) \mathcal{O}(t_4)  \rangle = \langle \textbf{T}^\dagger \mathcal{O}(t_1) \mathcal{O}(t_2) \mathcal{O}(t_3) \mathcal{O}(t_4) \textbf{T} \rangle^\ast
\end{equation}
Using time reversal \eqref{eqn:time_reversal_HO} and hermiticity of $\mathcal{O}$, we get
\begin{equation}
\langle  \mathcal{O}(t_1) \mathcal{O}(t_2) \mathcal{O}(t_3) \mathcal{O}(t_4)  \rangle = \langle  \mathcal{O}(-t_1) \mathcal{O}(-t_2) \mathcal{O}(-t_3) \mathcal{O}(-t_4) \rangle^\ast = \langle  \mathcal{O}(-t_4) \mathcal{O}(-t_3) \mathcal{O}(-t_2) \mathcal{O}(-t_1) \rangle 
\end{equation}
Let us choose $t_1 = t_4 = 0$ and $t_2=t_3 = t$. Then we get
\begin{equation}
\langle  \mathcal{O}(0) \mathcal{O}(t) \mathcal{O}(t) \mathcal{O}(0)  \rangle = \langle  \mathcal{O}(0) \mathcal{O}(-t) \mathcal{O}(-t) \mathcal{O}(0) \rangle \,.
\end{equation}
Note that the LHS is a SK correlation function for $t>0$ since it can be captured in SK contour\cite{Kamenev,Haehl}. But the RHS does not belong to SK correlation - it is rather an out-of-time-ordered (OTO) correlation function. (A review on generalised OTO correlation can be found in \cite{Haehl:2017qfl,Chaudhuri:2018ymp,Chaudhuri:2018ihk}.) Now, we impose thermality on the RHS via the Kubo-Martin-Sschwinger (KMS) \cite{Kubo:1957mj,Martin:1959jp} relation and find
\begin{equation}\label{eqn:time_reversal_thermality}
\langle  \mathcal{O}(0) \mathcal{O}(t) \mathcal{O}(t) \mathcal{O}(0)  \rangle = \langle  \mathcal{O}(-i\beta) \mathcal{O}(0) \mathcal{O}(-t) \mathcal{O}(-t) \rangle \,.
\end{equation}
Thus we find that two different SK correlation functions are related by both time-reversal invariance and thermality. When we integrate out the bath then bath correlations at the system time scale appears as parameters in the effective theory. Thus from \eqref{eqn:time_reversal_thermality}, we expect to get certain relation between corresponding parameters. This motivates the idea of why should there be a non-linear FDR under time reversal invariance and thermal nature of the bath.

Note that, for our system, there is no Onsager like relations at the level of two point correlations. The effective path integral \eqref{eqn:SK_effective_path_integral} describes only one particle with two parameters ($f^r,\gamma^r$) at quadratic order. Corresponding to these parameters there are two two-point bath correlations \cite{Chakrabarty:2019qcp} which are related by the KMS condition. Consequently, $f^r$ and $\gamma^r$ are related by the linear FDR \eqref{eqn:LinFDR}. There are no other parameters so that time reversal invariance would act and give non-trivial relation(s) among those parameters. Therefore, time reversal invariance acts trivially at the level of two point correlations for our system.

\subsection{Equal-time velocity four point function}
\label{subsec:vel_four_pt_sk}

We compute equal-time velocity four point function to study the implication of the non-linear FDR \eqref{eqn:nonLinFDR}.
The connected part of the equal-time velocity four point function is defined as
\begin{equation}\label{eqn:quartic_velocity_conn}
\langle \dot{q}^4(t) \rangle_c = \langle \dot{q}(t)^4 \rangle - 3 \langle \dot{q}^2(t) \rangle^2 \,.
\end{equation}
We compute it perturbatively from the path-integral \eqref{eqn:SK_effective_path_integral} using Feynman diagrams. As Feynman rules, we need the propagators and the vertex factors. The propagators are evaluated in appendix \ref{app:SK_prop} and the vertex factors for $q_d^4$ and $\dot{q} q_d^3$ vertices are given by $(-\zeta_\eta^r)$ and $(i 4!\zeta_\gamma^r)$ respectively.

The SK diagrams contributing to $\langle \dot{q}^4(t) \rangle_c$ at leading order, are drawn in figure \ref{fig:leading_contribution_v4}. We find that both $\zeta_\eta^r$ and $\zeta_\gamma^r$ contribute at the tree level. One can show that all loop diagrams consist of only sub-leading Grassmann loops, provided the propagator $G^d$ in \eqref{eqn:SK_propagatorD} is zero. Hence, tree level contribution to \eqref{eqn:quartic_velocity_conn} is the most significant contribution.

\begin{figure}
\begin{center}
\begin{tikzpicture}
\node at (-4,0) {$\left\langle (\dot{q}(t))^4 \right\rangle_c\, = $};
\begin{scope}[scale=0.5]
\draw[ultra thick, blue] (-2,2) -- (-1,1);
\draw[ultra thick, blue,dashed] (-1,1) -- (0,0);
\draw[ultra thick, blue] (2,2) -- (1,1);
\draw[ultra thick, blue,dashed] (1,1) -- (0,0); 
\draw[ultra thick, blue] (2,-2) -- (1,-1);
\draw[ultra thick, blue,dashed] (1,-1) -- (0,0); 
\draw[ultra thick, blue] (-2,-2) -- (-1,-1);
\draw[ultra thick, blue,dashed] (-1,-1) -- (0,0); 
\node at (-3.5,0) {$(-\zeta_\eta^r) \, \times $};

\node at (-1.8,2.8) {$\dot{q}(t)$};
\node at (1.8,2.8) {$\dot{q}(t)$};
\node at (-1.8,-2.8) {$\dot{q}(t)$};
\node at (1.8,-2.8) {$\dot{q}(t)$};

\node at (1,0) {$t'$};
\end{scope}

\begin{scope}[scale=0.5,shift={(10,0)}]
\draw[ultra thick, blue] (-2,2) -- (0,0);
\draw[ultra thick, blue] (2,2) -- (1,1);
\draw[ultra thick, blue,dashed] (1,1) -- (0,0); 
\draw[ultra thick, blue] (2,-2) -- (1,-1);
\draw[ultra thick, blue,dashed] (1,-1) -- (0,0); 
\draw[ultra thick, blue] (-2,-2) -- (-1,-1);
\draw[ultra thick, blue,dashed] (-1,-1) -- (0,0); 
\node at (-5,0) {$ + \, \, \, (i 4!\zeta_\gamma^r) \, \times  $};

\node at (-1.8,2.8) {$\dot{q}(t)$};
\node at (1.8,2.8) {$\dot{q}(t)$};
\node at (-1.8,-2.8) {$\dot{q}(t)$};
\node at (1.8,-2.8) {$\dot{q}(t)$};

\node at (1,0) {$t'$};
\end{scope}
\end{tikzpicture}
\end{center}
\caption{Leading order contribution to $ \langle \dot{q}^4(t)\rangle _c$: The vertex factors of the SK diagrams are $(-\zeta_\eta^r)$ and $(4! i\zeta_\gamma^r)$ respectively. We assume an out-going time convention which means time flows from $t'$ to $t$. In this convention, all of the propagators (except the Keldysh propagator on the right figure) are retarded propagators.}
\label{fig:leading_contribution_v4}
\end{figure}
To evaluate the SK diagrams, we choose an 'out-going time' convention for the diagrams in figure \ref{fig:leading_contribution_v4}. According to this convention, all propagators (except the Keldysh propagator) are retarded.
The integral representation of the SK diagrams in figure \ref{fig:leading_contribution_v4} is given by
\begin{equation}\label{eqn:v4}
\begin{split}
\langle \dot{q}^4(t) \rangle_c = & \, -\zeta_\eta^r  \int_{-\infty}^{\infty} dt' \langle \dot{q}(t) q_d(t') \rangle^4  
+(4! i \zeta_\gamma^r) \int_{-\infty}^{\infty} dt'  \langle \dot{q}(t) q_d(t') \rangle^3  \langle \dot{q}(t')\dot{q}(t) \rangle \,.
\end{split}
\end{equation}
Substituting \eqref{eqn:SK_propagatorK} and \eqref{eqn:SK_propagatorR} we evaluate the integrals. Then we use the linear and non-linear FDRs \eqref{eqn:FDRs} to get
\begin{equation}\label{eqn:v4_analytic}
\langle \dot{q}^4(t) \rangle_c = 
-\frac{\zeta^r_\eta}{4\gamma^r}  -4! \zeta_\gamma^r \frac{f^r}{2\gamma^r} \frac{1}{4\gamma^r} 
= \frac{\zeta^r_\eta}{4\gamma^r} \,. 
\end{equation}
Note that the above quantity is a negative number since $\zeta_\eta^r$ is negative as pointed out in the introduction. We define a quartic deviation (analogous to standard deviation) of velocity as
\begin{equation}\label{eqn:quartic_dev_vel}
v_{QD} \equiv - \vert \langle \dot{q}^4(t) \rangle_c \vert^{1/4} \,
\end{equation}
and match with the numeric in figure \ref{fig:v4fn}.

\section{A numerical analysis of velocity correlations}
\label{sec:numerical_analysis}

\subsection{Preparation of noise}
\label{subsec:preperation_noise}

\begin{figure}
\centering
\includegraphics[width=8cm, height=6cm] {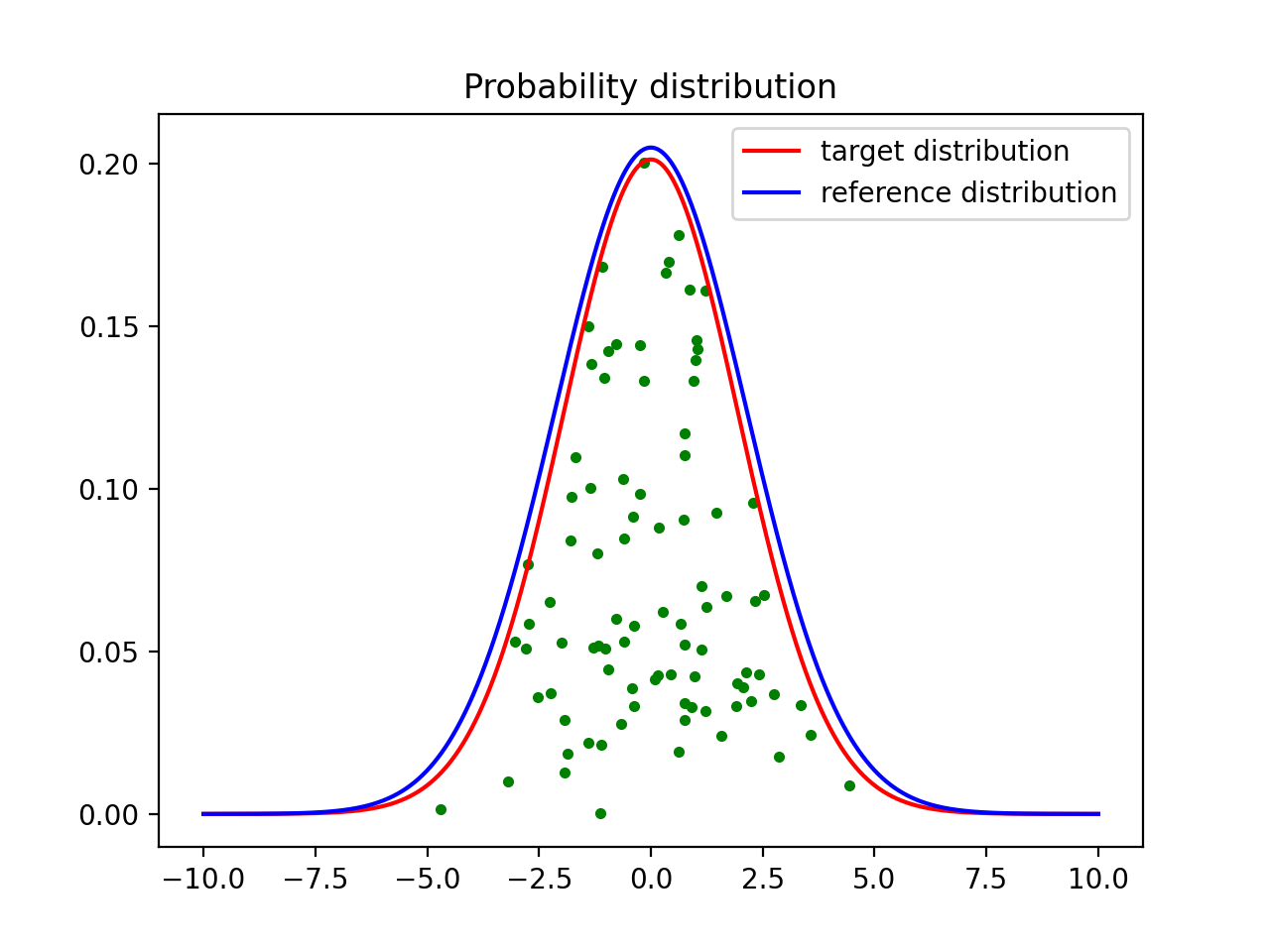}
\caption{The target distribution $P$ is found from the known reference distribution $P_g$. $P_g$ is chosen to be a scaled Gaussian distribution whose exact form is not relevant; it only has to contain the full target distribution. The target distribution $P(\eta) \sim e^{-\delta t \left( \frac{f}{2} \eta^2 + \frac{\zeta_\eta}{4!} \eta^4 \right) }$. The green dots generated from the rejection sampling, plot $u \times P_g$. Note that these dots do not go out of the red curve.}

\label{fig:noise_gen}
\end{figure}
We prepare the non-Gaussian noise distribution $P(\eta)$ in \eqref{eqn:nonG_noise} from a reference Gaussian distribution by a method called \textbf{rejection sampling}\footnote{I thank Anugu Sumith Reddy for bringing the rejection sampling method to my notice.}\cite{vonNeumann1951} originally proposed by von Neumann. (One could use different numerical methods such as Metropolis Hastings \cite{doi:10.1080/00031305.1995.10476177}). According to rejection sampling, we choose a reference noise distribution (Gaussian) function $P_g(\eta)$. The reference distribution $P_g(\eta)$ need not be normalised and should be such that it contains the full target distribution $P(\eta)$ within, as shown in figure \ref{fig:noise_gen}.
Then we choose a random number $u$ between $[0,1]$.  If $u<\frac{P(\eta)}{P_g(\eta)}$ for a generated $\eta$ from $P_g$, we keep that $\eta$ for $P$; else we discard. By repeating this process many times we can generate $P$. Out of 100 iteration figure \ref{fig:noise_gen} plots those data which follow the above inequality. An infinite number of iteration should populate the entire target distribution.

\begin{figure}[ht]
\centering
\begin{tikzpicture}
\begin{scope}[shift={(0,0)}]
\pgftext{\includegraphics[width=150pt]{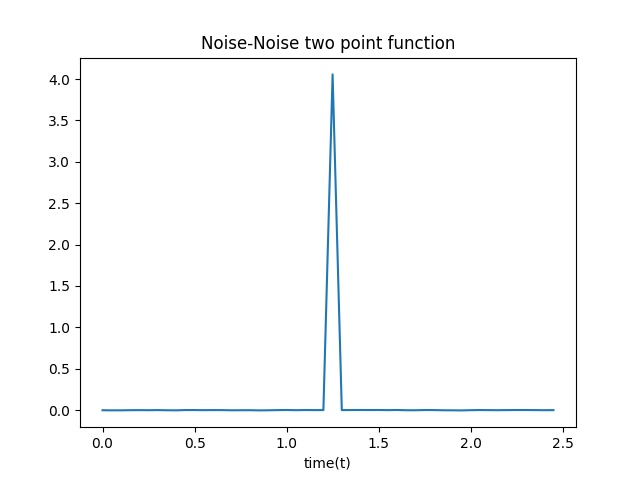}} at (0,0);
\node at (0,-2.2) {$(a)$};
\end{scope}
\begin{scope}[shift={(5,0)}]
\pgftext{\includegraphics[width=150pt]{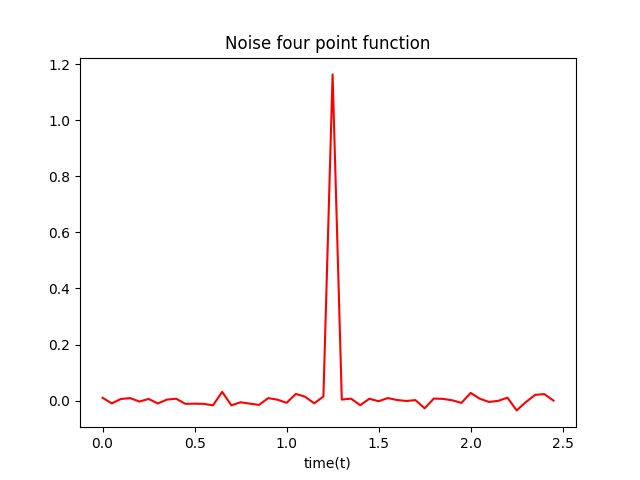}} at (0,0);
\node at (0,-2.2) {$(b)$};
\end{scope}
\begin{scope}[shift={(10,0)}]
\pgftext{\includegraphics[width=150pt]{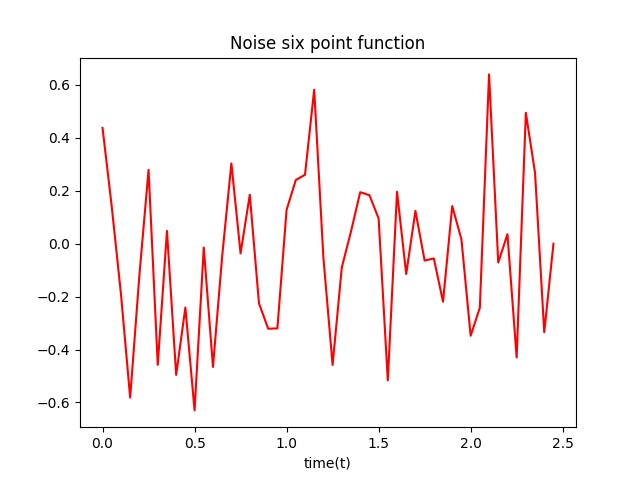}} at (0,0);
\node at (0,-2.2) {$(c)$};
\end{scope}
\end{tikzpicture}

\caption{We compute $\langle \eta(t)\eta(1.25)\rangle $, $\langle \eta(t)\eta(1.25)^3\rangle $ and $\langle \eta(t)\eta(1.25)^5\rangle$ in figure $(a),\,(b)$ and $(c)$ respectively. Here, the horizontal axis is in some unit of time. So, both $t, 1.25$ are numerical value in that unit of time. Since the noise does not have any dynamics, we get a peak (non-zero correlation) only when time of all noise variable are equal. The parameters for non-Gaussian noise distribution are chosen as $f=5.11$ and $\zeta_\eta = -0.086$ which are also the bare parameters discussed in section \ref{sec:renorm_parameter}. The units of these parameters are given in footnote \ref{footnote_units}. The six point function in $(c)$ is essentially indistinguishable from the fluctuation. The number of iteration for each graph is $10^7$.}
\label{fig:noise_corr}
\end{figure}

It is worthwhile to point out a subtlety in the target noise distribution $P(\eta)$. Derivation of the non-linear Langevin equation from microscopic theories \cite{Chakrabarty:2019aeu,Jana:2020vyx} finds that the actual numerical value of $\zeta_\eta$ is negative which makes $P(\eta)$ ill defined since it diverges as $\eta\rightarrow \infty$. To stabilise it, we add a small regulator such that the regulated distribution is given by
\begin{equation}\label{eqn:nonG_noise_R}
P_{\text{regulated}}(\eta) \sim e^{-\delta t \left( \frac{ f }{2} \eta^2 + \frac{\zeta_\eta}{4!} \eta^4 + \epsilon \eta^6 \right)} \,, \,\,\, 1>>\epsilon > 0 \,.
\end{equation}
The justification of the above regulator could be the following. All derivations of the Langevin equation consider only the leading order influence phase \cite{Feynman:1963fq} from the bath. The sub-leading correction should stabilise the distribution to make it a sensible theory.
We choose $\epsilon$ such that it nullifies any sixth cumulant generated during the noise preparation. We generate sufficient number ($10^8$) of $\eta$ using rejection sampling and plot two, four and six-point connected noise correlations. Since $\eta$ is a white noise,
 all of its correlations can have peak (non-zero values) when all $\eta$ are evaluated at the same time. So, we choose to vary one time-argument of each correlation, keeping the rest fixed at equal times as the following:
 \begin{equation}\label{eqn:noise_corr}
 \langle \eta(t) \eta(t') \rangle \,, \qquad  \left\langle \eta(t) (\eta(t'))^3 \right\rangle_c \qquad \text{and} \qquad \left\langle \eta(t) (\eta(t'))^5 \right\rangle_c \,.
 \end{equation}
 The connected four and six point correlations are defined as
 \begin{eqnarray}
\left\langle \eta(t) (\eta(t'))^3 \right\rangle_c &=&  \left\langle \eta(t) (\eta(t'))^3 \right\rangle - 3  \left\langle (\eta(t'))^2 \right\rangle \langle \eta(t) \eta(t') \rangle 
\\
\qquad \left\langle \eta(t) (\eta(t'))^5 \right\rangle_c &=& \qquad \left\langle \eta(t) (\eta(t'))^5 \right\rangle - 5  \left\langle (\eta(t'))^2 \right\rangle \left\langle \eta(t) (\eta(t'))^3 \right\rangle_c \nonumber
\\
&& - 10 \left\langle (\eta(t'))^4 \right\rangle_c \left\langle \eta(t) \eta(t') \right\rangle - 15 \left\langle (\eta(t'))^2 \right\rangle^2 \left\langle \eta(t) \eta(t') \right\rangle
 \end{eqnarray}
 where correlations are evaluated from the below explicit formula.
\begin{eqnarray}
&&
\langle \eta(t) (\eta(t'))^n \rangle = \frac{1}{N}\sum_{i=1}^N \eta_i(t) (\eta_i(t'))^n \,, \qquad N \text{ being the number of iteration.} 
\end{eqnarray}
We depict the above two point and connected four \& six point correlations in figure \ref{fig:noise_corr}. We find that two and four point functions show sharp peak while the six point function shows only fluctuation. This ensures that the noise distribution have insignificant sixth cumulant as required. Since we have successfully prepared the noise, we are ready to study the particle's dynamics numerically.
%
%

\subsubsection*{Comment on the regime of parameters:}
\label{subsubsec:comment_regime_parameter}

The bare parameters given in \eqref{eqn:nonLin_LangevinEqn} are used in the numerical computation whereas the renormalised ones are used in the analytics. The renormalised parameters are related to the bare parameters by an infinite series as written in \eqref{eqn:reln_bare_renorm_parameters}.
We evaluate till second order noise loop corrections in this series. This means our choice of bare parameters should be such that the convergence is fast enough to ensure matching of numeric and analytic data. We choose the bare parameters and the time step as
\begin{equation}\label{eqn:bare_parameters}
f =  5.11 \,,        \qquad        \zeta_\eta =  -0.086  \qquad \text{and}\qquad  \delta t = 0.05  
\end{equation}
respectively so that the perturbation parameter $\frac{\zeta_\eta}{f^2 \delta t}( = 0.066)$ in \eqref{eqn:reln_bare_renorm_parameters} makes  the second order term small (compared to the first order term). We also choose the numerical value of temperature to be $\frac{1}{\beta}=2$. The corresponding renormalised parameters\footnote{\label{footnote_units} We have not explicitly written the units of the parameters in the main text. If we assume the dimensions of $\dot{q}$ and $ t $ as [length][time]$^{-1}$ and [time] respectively, then the dimensions of $f$ and $\gamma$ are found to be [length]$^2$[time]$^{-3}$ and [time]$^{-1}$ respectively. Similarly, the dimensions of $\zeta_\eta,\,\zeta_\gamma$ are respectively [length]$^4$ [time]$^{-5}$, [length]$^2$[time]$^{-3}$. The dimension of $\beta$, using the FDRs, is given by [time]$^2$ [length]$^{-2}$ for unit mass and unit Boltzmann constant. But, most importantly, the perturbation parameter $\zeta_\eta/(f^2\delta t)$ is a dimensionless number.} are given by
\begin{equation}
f^r = 5.29 \,, \qquad \gamma^r = 1.31 \,, \qquad \zeta_\eta^r = -0.112 \,, \qquad \zeta_\gamma^r = 0.0046 \,\,.
\end{equation}

Let us now ask the following question. How much can we deviate from \eqref{eqn:bare_parameters} - the values of individual bare parameters, keeping the perturbation parameter at same order? If we make each parameter larger then the Euler-Maruyama error grows due to large values of $f$ and $\delta t$. Thus we will find deviation from analytic results. On the other hand, if we make each parameter smaller than in \eqref{eqn:bare_parameters}, we observe the following. We may get rid of the Euler-Maruyama error, but small $f$ and $\delta t$ bring large fluctuation in higher cumulants. For example, for $f = 1,\, \zeta_\eta = -0.001$ and $\delta t = 0.5$, $5\times 10^8$ number of ensemble is insufficient to tame the fluctuation and get a definitive plot for connected velocity four point function defined in \eqref{eqn:quartic_velocity_conn}. Thus we conclude that we should not make the parameters too large or too small. However, if one gets improvements on either side by better numerical methods, then the domain of validity of parameters can be made bigger.

\subsection{Velocity variance}
\label{subsec:velocity_var}

In this section, we numerically compute the velocity variance of the non-linear Langevin particle and show that its velocity variance matches to a high accuracy with that of linear Langevin particle at same temperature.

\begin{figure}
\centering

\includegraphics[width=9cm, height=7cm] {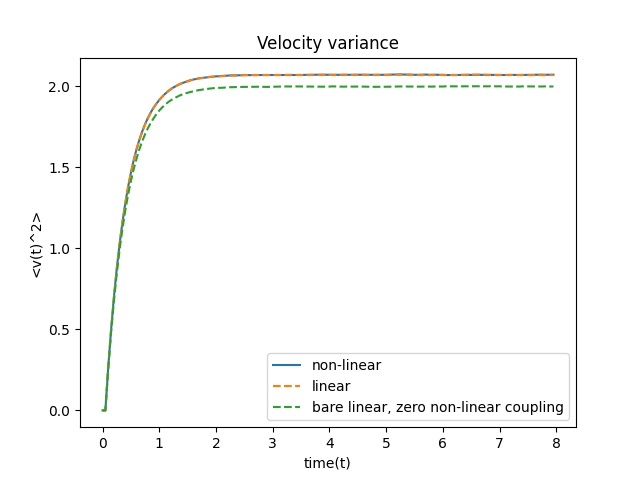}
   
\caption{The numerical value of bare parameters are given by $f=5.11,\,\gamma=1.31,\,\zeta_\eta=-0.086,\, \zeta_\gamma = 0.0042$ in respective units \ref{footnote_units}. The corresponding renormalised parameters are evaluated from \eqref{eqn:reln_bare_renorm_parameters} as $f^r = 5.29,\, \gamma^r = 1.32,\,\zeta_\eta^r = -0.112$ and $\zeta_\gamma^r = 0.0046 $ for $\delta t = 0.05$ . We assume both initial position and momentum as zero and find that velocity variance exactly matches in both linear and non-linear cases. Notice that saturation value does not occur exactly at temperature $\frac{1}{\beta}=\frac{f^r}{2\gamma^2} = 2$ due to the Euler-Maruyama error. We have chosen $f,\delta t$ large enough that the error is noticeable. However, since we are matching with linear Langevin result which also have identical error, the apparent deviation is irrelevant. The dashed line plots velocity variance with bare linear parameters but zero non-linear parameters. The number of iteration for each curve is $10^7$.}
\label{fig:vel_var}
\end{figure}

In order to study the dynamics of \eqref{eqn:nonLin_LangevinEqn}, we discretise it as the following.
\begin{equation}\label{eqn:discrete_nonLinLangevin}
\begin{split}
q(t+\delta t) =& \, q(t) + \delta t \, \dot{q} \,, \\
\dot{q}(t+\delta t) =& \, \dot{q}(t) - \delta t [\gamma+\zeta_{\gamma} \eta^2(t) ] \dot{q}(t) + f \delta t\, \eta(t) \,.
\end{split}
\end{equation}
The key point is that the parameters in the above equation are bare parameters. With these parameters we plot variance of velocity of the particle in figure \ref{fig:vel_var}. In this figure, we also simulate the linear Langevin equation with renormalised linear couplings ($f^r,\gamma^r$) and find that the velocity variances in both cases match. One might argue, discrepancy between them in figure \ref{fig:vel_var} are not observed due to \emph{weak} system-bath coupling and thus there is no significant distinction between bare and renormalised parameters. This is not correct. Had we used bare linear parameters ($f,\gamma$) and made $\zeta_\eta,\zeta_\gamma=0$ in \eqref{eqn:nonLin_LangevinEqn}, we would have got the dashed curve. So, the weak non-linear couplings do make a significant contribution.

\subsection{Velocity four point function and non-linear FDR}
\label{subsec:ckeck_nonFDR}

\begin{figure}
\centering
\includegraphics[width=9cm, height=5.5cm] {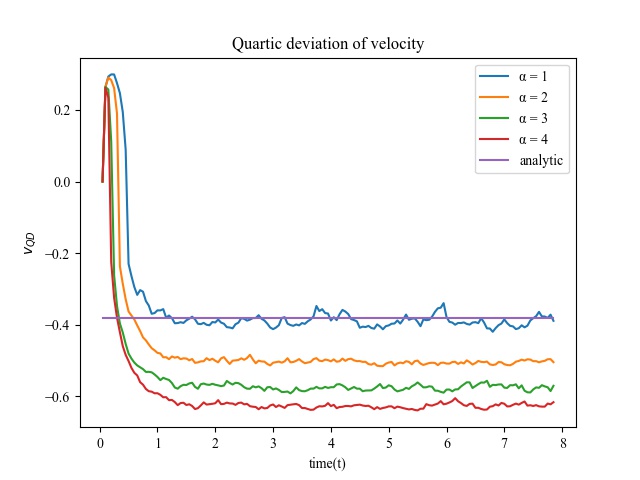}
   
\caption{We plot the quartic deviation $v_{QD}(t)\vert_{\text{numeric}}$ \eqref{eqn:def_frth_root_v4} for $\alpha=1,2,3,4$ as defined in \eqref{eqn:dev_FDR}. We also plot the analytic, thermal $v_{QD}$ as found in \eqref{eqn:v4_analytic} and find that $v_{QD}$ agrees well with $v_{QD}(t)\vert_{\text{numeric}}$ when $\alpha=1$, i.e., non-linear FDR is satisfied. The numerical value of the bare parameters for $\alpha=1$ are chosen as $f=5.11,\,\gamma=1.31,\,\zeta_\eta=-0.086,\, \zeta_\gamma = 0.0042$ in respective units. For $\alpha\neq 1$, only $\zeta_\gamma$s are different and are chosen as $0.0084,\, 0.0126,\, 0.0167$ to get $\alpha=2,\,3,\,4$ respectively. We find that fluctuation in $v_{QD}(t)\vert_{\text{numeric}}$ reduces as we make $\alpha$ bigger. So, we consider different number of iteration for each plot. The numbers of iteration are $5\times 10^7$ for $\alpha=1$, $3\times 10^7$ for $\alpha=2$ and $10^7$ for $\alpha=3,4$ each.}
\label{fig:v4fn}
\end{figure}

We find numerically the connected velocity four point function defined in \eqref{eqn:quartic_velocity_conn} for $5\times 10^7$ iterations. However, due to our choice of parameters, velocity four point function is comparable to typical fluctuation for the above number of iterations. We can reduce the fluctuation substantially by taking a fourth root of the velocity four point function as the following.
\begin{eqnarray}\label{eqn:def_frth_root_v4}
v_{QD}(t)\big\vert_\text{numeric} \equiv \begin{cases}
\left(\langle \dot{q}^4(t) \rangle_c\right)^{1/4} \,\,, \qquad \text{if} \qquad \langle \dot{q}^4(t) \rangle_c > 0 \\
-\left\vert \langle \dot{q}^4(t) \rangle_c \right\vert^{1/4} \,\,, \qquad \text{if} \qquad \langle \dot{q}^4(t) \rangle_c < 0
\end{cases}
\end{eqnarray}
The above definition for numerical analysis is different from \eqref{eqn:quartic_dev_vel} due to our prior ignorance about the sign of $\langle \dot{q}^4(t) \rangle_c $ in numerical analysis. We plot the above quantity in figure \ref{fig:v4fn} and find that the saturation value of the numerical data matches quite well with the analytic result when the non-linear FDR \eqref{eqn:nonLinFDR} is satisfied.

 In figure \ref{fig:v4fn}, there is a bump during the initial time evolution. To understand this feature, let us consider the discretised non-linear Langevin equation \eqref{eqn:discrete_nonLinLangevin} at $t=0$. Since the initial velocity is zero, there is no $\zeta_\gamma$ contribution at $t=0$. But we still have a $\zeta_\eta$ contribution from the noise which has a positive cumulant. Thus we get a positive bump at $t = \delta t$. After a few more time-steps $\zeta_\gamma$ term takes over since it has a larger negative contribution as pointed out in \eqref{eqn:v4}.

We also check how the quartic deviation behaves if we violate the non-linear FDR. We choose different values of the bare parameter $\zeta_\gamma$ such that the FDR differs by a factor of $\alpha$ defined as the following: 
\begin{equation}\label{eqn:dev_FDR}
{\zeta_\gamma^r}_\text{dev} \equiv \alpha \frac{-\zeta_\eta^r}{12} \,,
\end{equation}
where $\alpha=1$ defines the FDR. We find that the numeric quartic deviations differ from the analytically evaluated thermal quartic deviations \eqref{eqn:quartic_dev_vel} when $\alpha\neq 1$, as shown in figure \ref{fig:v4fn}. (Note that $\alpha=1$ for the analytic data in figure \ref{fig:v4fn} as corresponds to thermal quartic deviation.) The interpretation of this disagreement is the following. If the time reversal invariance of bath is broken ($\alpha \neq 1$), then we do not expect the particle to thermalise \cite{Berges:2000ur,2002cond.mat..2501M,Harris_2007,Parrondo_2009,Andrieux_2009,PhysRevLett.101.090602}. Rather, in our case, it reaches a steady state.

\section{Conclusion}
\label{sec:conclusion}

This work deals with systematic analysis of non-linear Langevin equation in presence of non-Gaussian noise. We show how the non-Gaussianity introduces the notion of renormalisation such that the bare/unphysical parameters are the parameters appearing in Langevin equation. On the other hand, the renormalised/physical parameters appear in the Schwinger-Keldysh effective action. We show, the bare and renormalised parameters are related by noise loop correction.

In the second half of this work, we do a numerical analysis to find the velocity variance and velocity four point function. We show that the velocity variance for both linear and non-linear Langevin dynamics, saturates to the temperature of the bath. We compute velocity four point function numerically and show that it agrees with thermal velocity four point function (calculated analytically) when non-linear FDR is satisfied. If we deviate from the non-linear FDR, the system does not thermalise but settles to a steady state.

\acknowledgments
I would like to thank Soumyadeep Chaudhuri for initial collaboration and guidance. I thank Arghya Das, Anupam Kundu, R. Loganayagam, Ajit Mehta, Avaya Pal, Archak Purkayastha, Anugu Sumith Reddy, Arnab Rudra, Prashant Singh, Rahul Singh and Akhil Sivakumar for useful suggestions.

\appendix

\section{Noise correlation functions in corrections of parameters}
\label{app:noise_corr}

We consider the path integral in \eqref{eqn:Ztailor_expnd} which evaluates correction to all parameters. To evaluate the loop diagrams we need the noise propagator as a Feynman rule. Since the noise does not have any dynamics, the propagator can only be Dirac-Delta function of time. But, in our numerical analysis, since the smallest available time scale is the time step $\delta t$, we write the following expression for noise propagator.
\begin{equation}
\begin{split}
\langle \eta(t_1) \eta(t_2) \rangle = 
\begin{cases} 
\frac{1}{f\delta t} \qquad \text{if} \qquad t_1 = t_2  \\
0 \qquad \text{else} \,.
\end{cases}
\end{split}
\end{equation}
The corresponding diagrammatic representation is given in figure \ref{fig:noise_prop}.
\begin{figure}
\centering
\begin{tikzpicture}

\draw (-1,0) node[below] {$t$} -- (1,0) node[below] {$t$};
\node at (1.5,0) {$ : $};
\node at (4,0) {$ \langle \eta(t) \eta(t) \rangle = \frac{1}{f\delta t}$};
\end{tikzpicture}
\caption{Diagrammatic representation of noise propagator}
\label{fig:noise_prop}
\end{figure}

\subsection*{Correction to \texorpdfstring{$q_d^2$}{} vertex:}
\label{subsec:corr_qd2}

The correction to $f$ are obtained from the noise two loop diagrams drawn in figure \ref{fig:qd2_corr}. The correction to $f$ is also pointed out below each diagram.
\begin{figure*}
\centering

\begin{tikzpicture}

\begin{scope}[shift={(-2,0)}]
\draw (0,0) circle (0.075);
\node at (0,0) {$\times$};
\node at (0,-1) {$-\frac{f^r}{2}$};
\end{scope}

\node at (-0.9,0) {$=$};

\begin{scope}[shift={(0,0)}]
\node at (0,0) {$\times$};
\node at (0,-1) {$-\frac{f}{2}$};
\end{scope}

\node at (1.3,0) {+};

\begin{scope}[shift={(2.4,0)}]
\draw (0,0.5) circle (0.5);
\node at (0,0) {$\times$};
\node at (0,-1) {$\frac{1}{4}\frac{\zeta_\eta}{f \delta t}$};
\end{scope}

\node at (3.7,0) {+};

\begin{scope}[shift={(5,0)}]
\draw (-0.5,0) -- (0.5,0);
\draw (0,0) circle (0.5);
\node at (0.5,0) {$\times$};
\node at (-0.5,0) {$\times$};
\node at (0,-1) {$-\frac{1}{12}\frac{\zeta_\eta^2}{f^3 \delta t^2}$};
\end{scope}

\node at (6.5,0) {+};

\begin{scope}[shift={(8,0)}]
\draw (-0.5,0) -- (0.5,0);
\draw (0.5,0.4) circle (0.4);
\draw (-0.5,0.4) circle (0.4);
\node at (0.5,0) {$\times$};
\node at (-0.5,0) {$\times$};
\node at (0,-1) {$-\frac{1}{8}\frac{\zeta_\eta^2}{f^3 \delta t^2}$};
\end{scope}

\node at (9.4,0) {+};

\begin{scope}[shift={(10.6,0)}]
\draw (0,0.3) circle (0.3);
\draw (0,0.9) circle (0.3);
\node at (0,0) {$\times$};
\node at (0,-1) {$-\frac{1}{8}\frac{\zeta_\eta^2}{f^3 \delta t^2}$};
\end{scope}

\end{tikzpicture}
\caption{Loop correction to $q_d^2$ vertex in SK path integral \eqref{eqn:Ztailor_expnd}. The corrected parameter is the renormalised parameter. }
\label{fig:qd2_corr}
\end{figure*}
Adding all the contributions we get
\begin{equation}
f^r = f - \frac{\zeta_\eta}{2  f \delta t} + \frac{2\zeta_\eta^2 }{3 f^3 (\delta t)^2} 
\end{equation}

\subsection*{Correction to \texorpdfstring{$q_d \dot{q}$}{} vertex:}
\label{subsec:corr_qd_v}

The diagrams which contribute to $q_d \dot{q}$ vertex are drawn in figure \ref{fig:qd_v_corr}.
\begin{figure*}
\centering

\begin{tikzpicture}

\begin{scope}[shift={(-2,0)}]
\draw (0,0) circle (0.075);
\node at (0,0) {$\times$};
\node at (0,-1) {$-i\gamma^r$};
\end{scope}

\node at (-0.9,0) {$=$};

\begin{scope}[shift={(0,0)}]
\node at (0,0) {$\times$};
\node at (0,-1) {$-i\gamma$};
\end{scope}

\node at (1.3,0) {+};

\begin{scope}[shift={(2.4,0)}]
\draw (0,0.5) circle (0.5);
\node at (0,0) {$\times$};
\node at (0,-1) {$\frac{-i\zeta_\gamma}{f \delta t}$};
\end{scope}

\node at (3.7,0) {+};

\begin{scope}[shift={(5,0)}]
\draw (0,0.3) circle (0.3);
\draw (0,0.9) circle (0.3);
\node at (0,0) {$\times$};
\node at (0,-1) {$\frac{i}{2}\frac{\zeta_\gamma \zeta_\eta}{f^3 \delta t^2}$};
\end{scope}

\end{tikzpicture}
\caption{Loop correction to $q_d \dot{q}$ vertex in SK path integral \eqref{eqn:Ztailor_expnd}.}
\label{fig:qd_v_corr}
\end{figure*}

Adding all contribution we get the following expression for $\gamma^r$.
\begin{equation}
\gamma^r = \gamma + \frac{\zeta_\gamma}{ f \delta t} - \frac{\zeta_\gamma \zeta_\eta}{2 f^3(\delta t)^2} \,.
\end{equation}

\subsection*{Correction to \texorpdfstring{$q_d^4$}{} vertex:}
\label{subsec:corr_qd4}

The correction to $q_d^4$ vertex gets contribution from the diagrams shown in figure \ref{fig:qd4_corr}.
\begin{figure*}
\centering

\begin{tikzpicture}

\begin{scope}[shift={(-2,0)}]
\draw (0,0) circle (0.075);
\node at (0,0) {$\times$};
\node at (0,-1) {$-\frac{1}{4!} \zeta_\eta^r$};
\end{scope}

\node at (-0.9,0) {$=$};

\begin{scope}[shift={(0,0)}]
\node at (0,0) {$\times$};
\node at (0,-1) {$-\frac{1}{4!} \zeta_\eta$};
\end{scope}

\node at (1.1,0) {+};

\begin{scope}[shift={(2.4,0)}]
\draw (0,0) circle (0.5);
\node at (-0.5,0) {$\times$};
\node at (0.5,0) {$\times$};
\node at (0,-1) {$\frac{3}{2\times 4!} \frac{\zeta_\eta^2}{f^2 \delta t}$};
\end{scope}

\node at (3.7,0) {+};

\begin{scope}[shift={(5.2,0)}]
\draw (-0.75,0) -- (0.5,0);
\draw (0.5,0.4) circle (0.4);
\node at (0.5,0) {$\times$};
\node at (-0.75,0) {$\times$};
\node at (0,-1) {$\frac{2}{4!}  \frac{\zeta_\eta^2}{f^2 \delta t}$};
\end{scope}

\node at (6.7,0) {+};

\begin{scope}[shift={(8.2,0)}]
\draw (0,0) circle (0.5);
\draw (0,0.8) circle (0.3);
\node at (-0.5,0) {$\times$};
\node at (0.5,0) {$\times$};
\node at (0,-1) {$ -\frac{3}{2\times4!} \frac{\zeta_\eta^3}{f^4 \delta t^2}$};
\end{scope}

\node at (9.5,0) {+};

\begin{scope}[shift={(11.2,0)}]
\draw (-0.4,0) circle (0.4);
\draw (0.4,0) circle (0.4);
\node at (-0.8,0) {$\times$};
\node at (0.8,0) {$\times$};
\node at (0,-1) {$-\frac{3}{4\times 4!}  \frac{\zeta_\eta^3}{f^4 \delta t^2}$};
\end{scope}


\node at (0.7,-3) {+};

\begin{scope}[shift={(2.2,-3)}]
\draw (-0.9,0) -- (0.5,0);
\draw (0.1,0) circle (0.4);
\node at (0.5,0) {$\times$};
\node at (-0.9,0) {$\times$};
\node at (0,-1) {$-\frac{2}{3\times4!}  \frac{\zeta_\eta^3}{f^4 \delta t^2}$};
\end{scope}

\node at (3.5,-3) {+};

\begin{scope}[shift={(5.1,-3)}]
\draw (-1,0) -- (0.5,0);
\draw (0.5,0.3) circle (0.3);
\draw (-0.3,0.3) circle (0.3);
\node at (0.5,0) {$\times$};
\node at (-1,0) {$\times$};
\node at (0,-1) {$-\frac{1}{4!}  \frac{\zeta_\eta^3}{f^4 \delta t^2}$};
\end{scope}

\node at (6.5,-3) {+};

\begin{scope}[shift={(8.2,-3)}]
\draw (-1,0) -- (0,0);
\draw (0,0.3) circle (0.3);
\draw (0,0.9) circle (0.3);
\node at (0,0) {$\times$};
\node at (-1,0) {$\times$};
\node at (0,-1) {$-\frac{1}{4!}  \frac{\zeta_\eta^3}{f^4 \delta t^2}$};
\end{scope}

\node at (9.3,-3) {+};

\begin{scope}[shift={(11.2,-3)}]
\draw (-1,0) -- (-0.3,0);
\draw (-1,0.3) circle (0.3);
\draw (0.1,0) circle (0.4);
\node at (-0.3,0) {$\times$};
\node at (-1,0) {$\times$};
\node at (0.5,0) {$\times$};
\node at (0,-1) {$-\frac{3}{4!}  \frac{\zeta_\eta^3}{f^4 \delta t^2}$};
\end{scope}


\node at (0.7,-6) {+};

\begin{scope}[shift={(2.2,-6)}]
\draw (0,0) circle (0.5);
\draw (-0.5,0) -- (0.5,0);
\node at (0.5,0) {$\times$};
\node at (-0.5,0) {$\times$};
\node at (0,-0.5) {$\times$};
\node at (0,-1) {$-\frac{3}{4!}  \frac{\zeta_\eta^3}{f^4 \delta t^2}$};
\end{scope}

\node at (3.5,-6) {+};

\begin{scope}[shift={(5.3,-6)}]
\draw (-1,0) -- (0.5,0);
\draw (0.5,0.3) circle (0.3);
\draw (-1,0.3) circle (0.3);
\node at (0.5,0) {$\times$};
\node at (-1,0) {$\times$};
\node at (-0.25,0) {$\times$};
\node at (0,-1) {$-\frac{3}{2\times 4!}  \frac{\zeta_\eta^3}{f^4 \delta t^2}$};
\end{scope}

\end{tikzpicture}
\caption{Loop correction to $q_d^4$ vertex in SK path integral \eqref{eqn:Ztailor_expnd}. The sixth 2-loop diagram contains two topologically distinct diagrams which have reflection symmetry.}
\label{fig:qd4_corr}
\end{figure*}
Combining all of these correction we get the renormalised $\zeta_\eta^r$ as
\begin{equation}
\zeta_\eta^r = \zeta_\eta - \frac{7}{2}\frac{\zeta_\eta^2}{ f^2\delta t } + \frac{149}{12} \frac{\zeta_\eta^3}{ f^4 (\delta t)^2} \,.
\end{equation}

\subsection*{Correction to \texorpdfstring{$q_d^3\dot{q}$}{} vertex:}
\label{subsec:corr_qd3v}

The correction to $q_d^3\dot{q}$ vertex gets contribution from the diagrams shown in figure \ref{fig:qd3v_corr}.
\begin{figure*}
\centering

\begin{tikzpicture}

\begin{scope}[shift={(-2,0)}]
\draw (0,0) circle (0.075);
\node at (0,0) {$\times$};
\node at (0,-1) {$i \zeta_\gamma^r$};
\end{scope}

\node at (-0.9,0) {$=$};

\begin{scope}[shift={(0,0)}]
\node at (0,0) {$\times$};
\node at (0,-1) {$i \zeta_\gamma$};
\end{scope}

\node at (1.1,0) {+};

\begin{scope}[shift={(2.4,0)}]
\draw (0,0) circle (0.5);
\node at (-0.5,0) {$\times$};
\node at (0.5,0) {$\times$};
\node at (0,-1) {$-\frac{i}{2} \frac{\zeta_\gamma \zeta_\eta}{f^2 \delta t}$};
\end{scope}

\node at (3.7,0) {+};

\begin{scope}[shift={(5.2,0)}]
\draw (-0.75,0) -- (0.5,0);
\draw (0.5,0.4) circle (0.4);
\node at (0.5,0) {$\times$};
\node at (-0.75,0) {$\times$};
\node at (0,-1) {$-i \frac{\zeta_\gamma \zeta_\eta}{f^2 \delta t}$};
\end{scope}

\node at (6.7,0) {+};

\begin{scope}[shift={(8.2,0)}]
\draw (0,0) circle (0.5);
\draw (0,0.8) circle (0.3);
\node at (-0.5,0) {$\times$};
\node at (0.5,0) {$\times$};
\node at (0,-1) {$ \frac{i}{2} \frac{\zeta_\eta \zeta_\gamma^2}{f^4 \delta t^2}$};
\end{scope}

\node at (9.5,0) {+};

\begin{scope}[shift={(11.2,0)}]
\draw (-0.4,0) circle (0.4);
\draw (0.4,0) circle (0.4);
\node at (-0.8,0) {$\times$};
\node at (0.8,0) {$\times$};
\node at (0,-1) {$ \frac{i}{4} \frac{\zeta_\eta \zeta_\gamma^2}{f^4 \delta t^2} $};
\end{scope}


\node at (0.7,-3) {+};

\begin{scope}[shift={(2.2,-3)}]
\draw (-0.9,0) -- (0.5,0);
\draw (0.1,0) circle (0.4);
\node at (0.5,0) {$\times$};
\node at (-0.9,0) {$\times$};
\node at (0,-1) {$ \frac{i}{3} \frac{\zeta_\eta \zeta_\gamma^2}{f^4 \delta t^2}  $};
\end{scope}

\node at (3.5,-3) {+};

\begin{scope}[shift={(5.1,-3)}]
\draw (-1,0) -- (0.5,0);
\draw (0.5,0.3) circle (0.3);
\draw (-0.3,0.3) circle (0.3);
\node at (0.5,0) {$\times$};
\node at (-1,0) {$\times$};
\node at (0,-1) {$ \frac{i}{2} \frac{\zeta_\eta \zeta_\gamma^2}{f^4 \delta t^2} $};
\end{scope}

\node at (6.5,-3) {+};

\begin{scope}[shift={(8.2,-3)}]
\draw (-1,0) -- (0,0);
\draw (0,0.3) circle (0.3);
\draw (0,0.9) circle (0.3);
\node at (0,0) {$\times$};
\node at (-1,0) {$\times$};
\node at (0,-1) {$ \frac{i}{2} \frac{\zeta_\eta \zeta_\gamma^2}{f^4 \delta t^2} $};
\end{scope}

\node at (9.3,-3) {+};

\begin{scope}[shift={(11.2,-3)}]
\draw (-1,0) -- (-0.3,0);
\draw (-1,0.3) circle (0.3);
\draw (0.1,0) circle (0.4);
\node at (-0.3,0) {$\times$};
\node at (-1,0) {$\times$};
\node at (0.5,0) {$\times$};
\node at (0,-1) {$ \frac{i}{2} \frac{\zeta_\eta \zeta_\gamma^2}{f^4 \delta t^2} $};
\end{scope}


\node at (0.7,-6) {+};

\begin{scope}[shift={(2.2,-6)}]
\draw (0,0) circle (0.5);
\draw (-0.5,0) -- (0.5,0);
\node at (0.5,0) {$\times$};
\node at (-0.5,0) {$\times$};
\node at (0,-0.5) {$\times$};
\node at (0,-1) {$ \frac{i}{2} \frac{\zeta_\eta \zeta_\gamma^2}{f^4 \delta t^2} $};
\end{scope}

\node at (3.5,-6) {+};

\begin{scope}[shift={(5.3,-6)}]
\draw (-1,0) -- (0.5,0);
\draw (0.5,0.3) circle (0.3);
\draw (-1,0.3) circle (0.3);
\node at (0.5,0) {$\times$};
\node at (-1,0) {$\times$};
\node at (-0.25,0) {$\times$};
\node at (0,-1) {$ \frac{i}{4} \frac{\zeta_\eta \zeta_\gamma^2}{f^4 \delta t^2} $};
\end{scope}

\end{tikzpicture}
\caption{Loop correction to $q_d^3\dot{q}$ vertex in SK path integral \eqref{eqn:Ztailor_expnd}. The sixth 2-loop diagram contains two topologically distinct diagrams which have reflection symmetry, as done in figure \ref{fig:qd4_corr}.}
\label{fig:qd3v_corr}
\end{figure*}
Combining all of these correction we get the renormalised $\zeta_\gamma^r$ as
\begin{equation}
\zeta_\gamma^r = \zeta_\gamma - \frac{3\zeta_\eta\zeta_\gamma}{2 f^2\delta t} + \frac{10}{3} \frac{\zeta_\eta^2\zeta_\gamma}{ f^4 (\delta t)^2}  \,.
\end{equation}

\section{Functional derivation of Schwinger-Keldysh propagators}
\label{app:SK_prop}

The propagators are obtained by considering the quadratic part of path-integral \eqref{eqn:SK_effective_path_integral}. Then we introduce sources corresponding to all variables. The propagators for a generic initial condition is hard to compute. However, we can compute the thermal propagators. We shift the variables and finally take functional derivative of the path integral w.r.t. the sources and obtain the thermal propagators.
The quadratic path integral with the sources is given by
\begin{equation}\label{eqn:Z_source}
\begin{split}
\mathcal{Z}_0 [J_{\dot{q}}, J_d, J_c, J_{\bar{c}} ] &=
\int \mathcal{D}\dot{q}\, \mathcal{D}q_d\, \mathcal{D}\bar{c} \, \mathcal{D}c \,\,  \exp \Big[ - \int dt \left\{  J_{\dot{q}}\, q_d + J_d\, \dot{q} + J_c \,c + \bar{c} \, J_{\bar{c}} \right\} \Big]\\
& 
\times \exp\Bigg[-\int dt \Bigg\{ i q_d \ddot{q} + i\gamma^r q_d \dot{q} + \frac{ f^r}{2} q_d^2 + i \bar{c}\left( \frac{d}{dt} + \gamma  \right)c \Bigg] \,.
\end{split}
\end{equation}
Here $J_{\dot{q}} ,\, J_d,\, J_c $ and $J_{\bar{c}}$ are sources of $q_d, \, \dot{q}, \, c$ and $\bar{c}$ respectively. It may seem that $q_d$ and $\dot{q}$ sources are flipped. But, it is a convention followed in SK literature. 

The path integral in Fourier space
\begin{equation}\label{eqn:Z_source_fourier}
\begin{split}
\mathcal{Z}_0 
= &
\int \mathcal{D}\dot{q}\, \mathcal{D}q_d\, \mathcal{D}\bar{c} \, \mathcal{D}c 
\times  e^{- \int \frac{d\omega}{2\pi} \left\{  \tilde{J}_{\dot{q}}(-\omega)\, \tilde{q}_d(\omega) + \tilde{J}_d(-\omega)\, \tilde{\dot{q}}(\omega) + \tilde{J}_c(-\omega) \,\tilde{c}(\omega) + \tilde{\bar{c}}(-\omega) \, \tilde{J}_{\bar{c}}(\omega) \right\} } \\
& 
\qquad \qquad \qquad  \times e^{ -\int \frac{d\omega}{2\pi} \left\{ \tilde{q}_d(-\omega) \left(\omega+i\gamma^r \right) \tilde{\dot{q}} + \frac{ f^r}{2} \tilde{q}_d(-\omega) \tilde{q}_d(\omega) + \tilde{\bar{c}} (-\omega) \left( \omega + i \gamma  \right) \tilde{c} (\omega) \right\}} \,.
\end{split}
\end{equation}
We give a shift to the variables as the following such that the variables and their corresponding sources are decoupled.\begin{eqnarray}
\tilde{q}_d'(\omega) &=& \tilde{q}_d(\omega) + \frac{\tilde{J_d}(\omega)}{-\omega+i\gamma^r} \,, \\
\tilde{\dot{q}}'(\omega) &=& \tilde{\dot{q}}(\omega) - \frac{f^r\, \tilde{J}_d(\omega)}{(-\omega+i\gamma^r)(\omega+i\gamma^r)} + \frac{\tilde{J}_{\dot{q}}(\omega)}{\omega + i\gamma^r} \,, 
\end{eqnarray}
\begin{eqnarray}
\tilde{c}'(\omega) =  \tilde{c}(\omega) + \frac{\tilde{J}_{\bar{c}}(\omega) }{\omega + i\gamma} \,, \qquad \tilde{\bar{c}}'(\omega) = \tilde{\bar{c}}(\omega) + \frac{\tilde{J}_c(\omega)}{-\omega+i\gamma}   \,.
\end{eqnarray}
The corresponding path-integral is given by
\begin{equation}\label{eqn:Zzero_source_fourier}
\begin{split}
\mathcal{Z}_0
= &
\int \mathcal{D}\dot{q}' \mathcal{D}q'_d \mathcal{D}\bar{c}'  \mathcal{D}c' \, \, e^{\int \frac{d\omega}{2\pi} \left\{ \frac{f^r}{2} \frac{ \tilde{J}_d(-\omega)\,  \tilde{J}_d(\omega)}{\omega^2+(\gamma^r)^2} - \frac{ \tilde{J}_{\dot{q}}(-\omega)\, \tilde{J}_d(\omega)}{\omega - i\gamma^r} + \frac{\tilde{J}_{c}(-\omega) \tilde{J}_{\bar{c}}(\omega)}{\omega+i\gamma} \right\} } \\
& \qquad \qquad \times
e^{- \int \frac{d\omega}{2\pi} \left\{ \frac{f^r}{2} \tilde{q}'_d(-\omega) \tilde{q}'_d(\omega) + (\omega+i\gamma^r) \tilde{q}'_d(-\omega) \tilde{\dot{q}}'(\omega) + (\omega+i\gamma)\tilde{\bar{c}}'(-\omega)\tilde{c}'(\omega) \right\}  }  \,.
\end{split}
\end{equation}
The first line in the above expression consists of only sources and the second line does not involve any source. Thus, integrating out the second line gives a constant which can be absorbed into the definition of $\mathcal{Z}_{0}$. Thus we find
\begin{equation}
\mathcal{Z}_0 = \exp \left[ \int \frac{d\omega}{2\pi} \left\{ \frac{f^r}{2} \frac{ \tilde{J}_d(-\omega)\,  \tilde{J}_d(\omega)}{\omega^2+(\gamma^r)^2} - \frac{ \tilde{J}_{\dot{q}}(-\omega)\, \tilde{J}_d(\omega)}{\omega - i\gamma^r} + \frac{\tilde{J}_{c}(-\omega) \tilde{J}_{\bar{c}}(\omega)}{\omega+i\gamma} \right\} \right] \,.
\end{equation}
Fourier transforming back to real time, we get
\begin{equation}
\mathcal{Z}_0 = \exp \left[ \int dt_1 dt_2 \int  \frac{d\omega}{2\pi} e^{-i\omega (t_2-t_1)} \left\{ \frac{f^r}{2} \frac{ J_d(t_2)\, J_d(t_1)}{\omega^2+(\gamma^r)^2} - \frac{ J_{\dot{q}}(t_2)\, J_d(t_1)}{\omega - i\gamma^r} + \frac{J_{c}(t_2) J_{\bar{c}}(t_1)}{\omega+i\gamma} \right\} \right]
\end{equation}
The propagators are obtained by taking functional derivative of $\mathcal{Z}_0$ w.r.t. the sources. The SK propagators are accordingly given by
\begin{eqnarray}
\langle \dot{q}(t_2) \dot{q}(t_1) \rangle &=& (-1)^2\frac{\partial^2 \mathcal{Z}_0}{\partial {J_q}(t_2) \,\partial {J_d}(t_1)}\Big\vert_{J_{\dot{q}}, J_d, J_c, J_{\bar{c}} = 0} =  \int \frac{d\omega}{2\pi}\, \frac{ f^r \, e^{-i\omega (t_2-t_1)} }{\omega^2+(\gamma^r)^2} \,,\\
\langle  q_d(t_2) \dot{q}(t_1) \rangle &=& (-1)^2\frac{\partial^2 \mathcal{Z}_0}{\partial J_{\dot{q}} (t_2)\, \partial J_d(t_1)} \Big\vert_{J_{\dot{q}}, J_d, J_c, J_{\bar{c}} = 0}  = - \int \frac{d\omega}{2\pi} \, \frac{ e^{-i\omega (t_2-t_1)} }{\omega-i\gamma^r} \,,\\
\langle \dot{q}(t_2) q_d(t_1) \rangle &=& (-1)^2 \frac{\partial^2 \mathcal{Z}_0}{ \partial J_d(t_2)\, \partial J_{\dot{q}} (t_1)} \Big\vert_{J_{\dot{q}}, J_d, J_c, J_{\bar{c}} = 0}  =  \int \frac{d\omega}{2\pi} \, \frac{ e^{-i\omega (t_2-t_1)} }{\omega+i\gamma^r} \,,\\
\langle q_d(t_2) q_d(t_1) \rangle &=& (-1)^2 \frac{\partial^2 \mathcal{Z}_0}{ \partial J_{\dot{q}}(t_2)\, \partial J_{\dot{q}} (t_1)} \Big\vert_{J_{\dot{q}}, J_d, J_c, J_{\bar{c}} = 0}  = 0 \,.
\end{eqnarray}
The first three propagators are called the Keldysh, advanced and retarded propagator respectively.
Performing the $\omega$ integral we get
\begin{eqnarray}\label{eqn:SK_prop_time}
\langle \dot{q}(t_2) \dot{q}(t_1) \rangle &=& \frac{f^r}{2\gamma^r} e^{-\gamma^r|t_2-t_1|}  \,, \\
\langle q_d(t_2) \dot{q}(t_1) \rangle &=& -i\, \Theta(t_1-t_2) e^{-\gamma^r (t_1-t_2)} \,,\\
\langle \dot{q}(t_2) q_d(t_1) \rangle &=&  -i \, \Theta(t_2-t_1) e^{-\gamma^r (t_2-t_1)} \,, \\
\langle q_d(t_2) q_d(t_1) \rangle &=&  0 \,.
\end{eqnarray}
We implement the above propagators in computation of real time correlation function in  \S\ref{subsec:skProp} and in \S\ref{subsec:vel_four_pt_sk}. Similarly the Grassmannian propagators are obtained as the following.
\begin{eqnarray}
\langle  \bar{c}(t_2) \, c(t_1) \rangle &=& (-1)^2\frac{\partial^2 \mathcal{Z}_0}{\partial J_{\bar{c}}(t_2) \,\partial J_c(t_1)}\Big\vert_{J_{\dot{q}}, J_d, J_c, J_{\bar{c}} = 0} \nonumber\\
&=&  -\int \frac{d\omega}{2\pi}\, \frac{ e^{-i\omega (t_2-t_1)} }{\omega-i\gamma}  \,.
\end{eqnarray}

\subsection*{Sub-leading correction to \texorpdfstring{$f^r,\zeta_\eta^r$}{fr} due to \texorpdfstring{$\bar{c}$}-\texorpdfstring{$c$}{}  loop:}
\label{app:corr_f_due_to_ccbar}

The correction to $f^r$ is obtained from a Grassmannian loop integral given by
\begin{eqnarray}
(f^r)' &=& f^r + (-1) (2i\zeta_\gamma)(-1) \int \frac{d\omega}{2\pi}  \frac{1}{\omega -i\gamma}  \nonumber \\
&=& f^r - \zeta_\gamma \,.
\end{eqnarray}
The first $(-1)$ appears due to the Grassmann loop, the factor $(2i\gamma)$ is the Feynman rules obtained from the effective path-integral \eqref{eqn:SK_effective_path_integral}. In our case $f^r = 5.29$ and $\zeta_\gamma = 0.0042$. So, the correction is smaller than the second order noise loop correction, thus can be ignored.

In a similar way we obtain the correction to $\zeta_\gamma^r$, given by
\begin{eqnarray}
(\zeta_\eta^r)' &=& \zeta_\eta^r + (-1) (2i\zeta_\gamma)^2 \int \frac{d\omega}{2\pi}  \frac{-1}{(\omega -i\gamma)(\omega+i\gamma)} \nonumber \\
&=& \zeta_\eta^r - \frac{2\zeta_\gamma^2}{\gamma} \,.
\end{eqnarray}
The correction is again sufficiently small. Thus we conclude that the Grassmannian correction, for the domain of our parameters, does not significantly contribute to correlation function.

\addcontentsline{toc}{section}{References}
\bibliographystyle{utphys} 
\bibliography{nonLinLangevin}

\providecommand{\href}[2]{#2}\begingroup\raggedright\begin{thebibliography}{10}

\bibitem{Chakrabarty:2019qcp}
B.~Chakrabarty and S.~Chaudhuri, ``{Out of time ordered effective dynamics of a
  quartic oscillator},''
  \href{http://dx.doi.org/10.21468/SciPostPhys.7.1.013}{{\em SciPost Phys.}
  {\bfseries 7} (2019) 013}, \href{http://arxiv.org/abs/1905.08307}{{\ttfamily
  arXiv:1905.08307 [hep-th]}}.

\bibitem{Caldeira1982PathIA}
A.~Caldeira and A.~J. Leggett, ``Path integral approach to quantum brownian
  motion,'' {\em Physica A-statistical Mechanics and Its Applications}
  {\bfseries 121} (1982) 587--616.

\bibitem{Chakrabarty:2018dov}
B.~Chakrabarty, S.~Chaudhuri, and R.~Loganayagam, ``{Out of Time Ordered
  Quantum Dissipation},'' \href{http://dx.doi.org/10.1007/JHEP07(2019)102}{{\em
  JHEP} {\bfseries 07} (2019) 102},
  \href{http://arxiv.org/abs/1811.01513}{{\ttfamily arXiv:1811.01513
  [cond-mat.stat-mech]}}.

\bibitem{Schwinger:1960qe}
J.~S. Schwinger, ``{Brownian motion of a quantum oscillator},''
\href{http://dx.doi.org/10.1063/1.1703727}{{\em J. Math. Phys.} {\bfseries 2}
  (1961) 407--432}.

\bibitem{Keldysh:1964ud}
L.~V. Keldysh, ``{Diagram technique for nonequilibrium processes},'' {\em Zh.
  Eksp. Teor. Fiz.} {\bfseries 47} (1964) 1515--1527.
[Sov. Phys. JETP20,1018(1965)].

\bibitem{Calzetta:1986ey}
E.~Calzetta and B.~L. Hu, ``{Closed Time Path Functional Formalism in Curved
  Space-Time: Application to Cosmological Back Reaction Problems},''
\href{http://dx.doi.org/10.1103/PhysRevD.35.495}{{\em Phys. Rev.} {\bfseries
  D35} (1987) 495}.

\bibitem{Kamenev}
A.~Kamenev, {\em Field Theory of Non-Equilibrium Systems}.
\newblock Cambridge University Press, Cambridge, 1985.
\newblock
  \url{http://www.cambridge.org/us/academic/subjects/physics/condensed-matter-physics-nanoscience-and-mesoscopic-physics/field-theory-non-equilibrium-systems?format=HB&isbn=9780521760829}.

\bibitem{Chakrabarty:2019aeu}
B.~Chakrabarty, J.~Chakravarty, S.~Chaudhuri, C.~Jana, R.~Loganayagam, and
  A.~Sivakumar, ``{Nonlinear Langevin dynamics via holography},''
  \href{http://dx.doi.org/10.1007/JHEP01(2020)165}{{\em JHEP} {\bfseries 01}
  (2020) 165}, \href{http://arxiv.org/abs/1906.07762}{{\ttfamily
  arXiv:1906.07762 [hep-th]}}.

\bibitem{Jana:2020vyx}
C.~Jana, R.~Loganayagam, and M.~Rangamani, ``{Open quantum systems and
  Schwinger-Keldysh holograms},''
  \href{http://dx.doi.org/10.1007/JHEP07(2020)242}{{\em JHEP} {\bfseries 07}
  (2020) 242}, \href{http://arxiv.org/abs/2004.02888}{{\ttfamily
  arXiv:2004.02888 [hep-th]}}.

\bibitem{Aharony:1999ti}
O.~Aharony, S.~S. Gubser, J.~M. Maldacena, H.~Ooguri, and Y.~Oz, ``{Large N
  field theories, string theory and gravity},''
  \href{http://dx.doi.org/10.1016/S0370-1573(99)00083-6}{{\em Phys. Rept.}
  {\bfseries 323} (2000) 183--386},
  \href{http://arxiv.org/abs/hep-th/9905111}{{\ttfamily arXiv:hep-th/9905111}}.

\bibitem{Glorioso:2018mmw}
P.~Glorioso, M.~Crossley, and H.~Liu, ``{A prescription for holographic
  Schwinger-Keldysh contour in non-equilibrium systems},''
  \href{http://arxiv.org/abs/1812.08785}{{\ttfamily arXiv:1812.08785
  [hep-th]}}.

\bibitem{Berges:2000ur}
J.~Berges and J.~Cox, ``{Thermalization of quantum fields from time reversal
  invariant evolution equations},''
  \href{http://dx.doi.org/10.1016/S0370-2693(01)01004-8}{{\em Phys. Lett. B}
  {\bfseries 517} (2001) 369--374},
  \href{http://arxiv.org/abs/hep-ph/0006160}{{\ttfamily arXiv:hep-ph/0006160}}.

\bibitem{Maldacena:2002vr}
J.~M. Maldacena, ``{Non-Gaussian features of primordial fluctuations in single
  field inflationary models},''
  \href{http://dx.doi.org/10.1088/1126-6708/2003/05/013}{{\em JHEP} {\bfseries
  05} (2003) 013}, \href{http://arxiv.org/abs/astro-ph/0210603}{{\ttfamily
  arXiv:astro-ph/0210603}}.

\bibitem{ALLEN198766}
T.~Allen, B.~Grinstein, and M.~B. Wise, ``Non-gaussian density perturbations in
  inflationary cosmologies,''
  \href{http://dx.doi.org/https://doi.org/10.1016/0370-2693(87)90343-1}{{\em
  Physics Letters B} {\bfseries 197} no.~1, (1987) 66--70}.
  \url{https://www.sciencedirect.com/science/article/pii/0370269387903431}.

\bibitem{Gangui:1993tt}
A.~Gangui, F.~Lucchin, S.~Matarrese, and S.~Mollerach, ``{The Three point
  correlation function of the cosmic microwave background in inflationary
  models},'' \href{http://dx.doi.org/10.1086/174421}{{\em Astrophys. J.}
  {\bfseries 430} (1994) 447--457},
  \href{http://arxiv.org/abs/astro-ph/9312033}{{\ttfamily
  arXiv:astro-ph/9312033}}.

\bibitem{Acquaviva:2002ud}
V.~Acquaviva, N.~Bartolo, S.~Matarrese, and A.~Riotto, ``{Second order
  cosmological perturbations from inflation},''
  \href{http://dx.doi.org/10.1016/S0550-3213(03)00550-9}{{\em Nucl. Phys. B}
  {\bfseries 667} (2003) 119--148},
  \href{http://arxiv.org/abs/astro-ph/0209156}{{\ttfamily
  arXiv:astro-ph/0209156}}.

\bibitem{PhysRevD.46.4232}
T.~Falk, R.~Rangarajan, and M.~Srednicki, ``Dependence of density perturbations
  on the coupling constant in a simple model of inflation,''
  \href{http://dx.doi.org/10.1103/PhysRevD.46.4232}{{\em Phys. Rev. D}
  {\bfseries 46} (Nov, 1992) 4232--4234}.
  \url{https://link.aps.org/doi/10.1103/PhysRevD.46.4232}.

\bibitem{10.1007/3-540-16452-9_6}
A.~A. Starobinsky, ``Stochastic de sitter (inflationary) stage in the early
  universe,'' in {\em Field Theory, Quantum Gravity and Strings}, H.~J. de~Vega
  and N.~S{\'a}nchez, eds., pp.~107--126.
\newblock Springer Berlin Heidelberg, Berlin, Heidelberg, 1986.

\bibitem{Goldberger:2004jt}
W.~D. Goldberger and I.~Z. Rothstein, ``{An Effective field theory of gravity
  for extended objects},''
  \href{http://dx.doi.org/10.1103/PhysRevD.73.104029}{{\em Phys. Rev. D}
  {\bfseries 73} (2006) 104029},
  \href{http://arxiv.org/abs/hep-th/0409156}{{\ttfamily arXiv:hep-th/0409156}}.

\bibitem{Galley:2009px}
C.~R. Galley and M.~Tiglio, ``{Radiation reaction and gravitational waves in
  the effective field theory approach},''
  \href{http://dx.doi.org/10.1103/PhysRevD.79.124027}{{\em Phys. Rev. D}
  {\bfseries 79} (2009) 124027},
  \href{http://arxiv.org/abs/0903.1122}{{\ttfamily arXiv:0903.1122 [gr-qc]}}.

\bibitem{Martin:1973zz}
P.~C. Martin, E.~D. Siggia, and H.~A. Rose, ``{Statistical Dynamics of
  Classical Systems},'' \href{http://dx.doi.org/10.1103/PhysRevA.8.423}{{\em
  Phys. Rev. A} {\bfseries 8} (1973) 423--437}.

\bibitem{DeDominicis:1977fw}
C.~De~Dominicis and L.~Peliti, ``{Field Theory Renormalization and Critical
  Dynamics Above t(c): Helium, Antiferromagnets and Liquid Gas Systems},''
  \href{http://dx.doi.org/10.1103/PhysRevB.18.353}{{\em Phys. Rev. B}
  {\bfseries 18} (1978) 353--376}.

\bibitem{article}
H.~Janssen, ``On a lagrangean for classical field dynamics and renormalization
  group calculations of dynamical critical properties,''
  \href{http://dx.doi.org/10.1007/BF01316547}{{\em Zeitschrift für Physik B
  Condensed Matter and Quanta} {\bfseries 23} (01, 1976) 377--380}.

\bibitem{Wang:1998wg}
E.~Wang and U.~W. Heinz, ``{A Generalized fluctuation dissipation theorem for
  nonlinear response functions},''
  \href{http://dx.doi.org/10.1103/PhysRevD.66.025008}{{\em Phys. Rev. D}
  {\bfseries 66} (2002) 025008},
  \href{http://arxiv.org/abs/hep-th/9809016}{{\ttfamily arXiv:hep-th/9809016}}.

\bibitem{vonNeumann1951}
J.~von Neumann, ``Various techniques used in connection with random digits,''
  in {\em Monte Carlo Method}, A.~S. Householder, G.~E. Forsythe, and H.~H.
  Germond, eds., vol.~12 of {\em National Bureau of Standards Applied
  Mathematics Series}, ch.~13, pp.~36--38.
\newblock US Government Printing Office, Washington, DC, 1951.

\bibitem{2002cond.mat..2501M}
C.~{Maes} and K.~{Netocny}, ``{Time-Reversal and Entropy},'' {\em arXiv
  e-prints} (Feb., 2002) cond--mat/0202501,
  \href{http://arxiv.org/abs/cond-mat/0202501}{{\ttfamily
  arXiv:cond-mat/0202501 [cond-mat.stat-mech]}}.

\bibitem{Harris_2007}
R.~J. Harris and G.~M. Schütz, ``Fluctuation theorems for stochastic
  dynamics,'' \href{http://dx.doi.org/10.1088/1742-5468/2007/07/p07020}{{\em
  Journal of Statistical Mechanics: Theory and Experiment} {\bfseries 2007}
  no.~07, (Jul, 2007) P07020--P07020}.
  \url{https://doi.org/10.1088/1742-5468/2007/07/p07020}.

\bibitem{Parrondo_2009}
J.~M.~R. Parrondo, C.~V. den Broeck, and R.~Kawai, ``Entropy production and the
  arrow of time,'' \href{http://dx.doi.org/10.1088/1367-2630/11/7/073008}{{\em
  New Journal of Physics} {\bfseries 11} no.~7, (Jul, 2009) 073008}.
  \url{https://doi.org/10.1088/1367-2630/11/7/073008}.

\bibitem{Andrieux_2009}
D.~Andrieux, P.~Gaspard, T.~Monnai, and S.~Tasaki, ``The fluctuation theorem
  for currents in open quantum systems,''
  \href{http://dx.doi.org/10.1088/1367-2630/11/4/043014}{{\em New Journal of
  Physics} {\bfseries 11} no.~4, (Apr, 2009) 043014}.
  \url{https://doi.org/10.1088/1367-2630/11/4/043014}.

\bibitem{PhysRevLett.101.090602}
E.~H. Feng and G.~E. Crooks, ``Length of time's arrow,''
  \href{http://dx.doi.org/10.1103/PhysRevLett.101.090602}{{\em Phys. Rev.
  Lett.} {\bfseries 101} (Aug, 2008) 090602}.
  \url{https://link.aps.org/doi/10.1103/PhysRevLett.101.090602}.

\bibitem{cite-key1}
I.~Oppenheim, ``Nonlinear nonequilibrium thermodynamics i. linear and nonlinear
  fluctuation-dissipation theorems,''
  \href{http://dx.doi.org/10.1007/BF02183157}{{\em Journal of Statistical
  Physics} {\bfseries 77} no.~5, (1994) 1109--1110}.
  \url{https://doi.org/10.1007/BF02183157}.

\bibitem{Stratonovich1992NonlinearNT}
R.~L. Stratonovich, ``Nonlinear nonequilibrium thermodynamics ii,''
\newblock 1992.

\bibitem{cite-key}
.~{\O}ksendal, B. K. (Bernt~Karsten), {\em Stochastic differential equations :
  an introduction with applications}.
\newblock Sixth edition. Berlin ; New York : Springer, {$[$}2003{$]$}
  {\copyright}2003, {$[$}2003{$]$}.
\newblock \url{https://search.library.wisc.edu/catalog/999949285502121}.

\bibitem{Chaudhuri:2020sc}
S.~Chaudhuri,
``{\, unpublished notes\, },''.

\bibitem{Onsager:1931jfa}
L.~Onsager, ``{Reciprocal Relations in Irreversible Processes. I.},''
  \href{http://dx.doi.org/10.1103/physrev.37.405}{{\em Phys. Rev.} {\bfseries
  37} no.~4, (1931) 405--426}.

\bibitem{Onsager:1931kxm}
L.~Onsager, ``{Reciprocal Relations in Irreversible Processes. II.},''
  \href{http://dx.doi.org/10.1103/physrev.38.2265}{{\em Phys. Rev.} {\bfseries
  38} no.~12, (1931) 2265--2279}.

\bibitem{RevModPhys.17.343}
H.~B.~G. Casimir, ``On onsager's principle of microscopic reversibility,''
  \href{http://dx.doi.org/10.1103/RevModPhys.17.343}{{\em Rev. Mod. Phys.}
  {\bfseries 17} (Apr, 1945) 343--350}.
  \url{https://link.aps.org/doi/10.1103/RevModPhys.17.343}.

\bibitem{Haehl}
F.~Haehl, R.~Loganayagam, and M.~Rangamani, ``Schwinger-keldysh formalism i:
  Brst symmetries and superspace,''
  \href{http://dx.doi.org/10.1007/JHEP06(2017)069}{{\em Journal of High Energy
  Physics} {\bfseries 2017} (10, 2016) }.

\bibitem{Haehl:2017qfl}
F.~M. Haehl, R.~Loganayagam, P.~Narayan, and M.~Rangamani, ``{Classification of
  out-of-time-order correlators},''
  \href{http://dx.doi.org/10.21468/SciPostPhys.6.1.001}{{\em SciPost Phys.}
  {\bfseries 6} no.~1, (2019) 001},
  \href{http://arxiv.org/abs/1701.02820}{{\ttfamily arXiv:1701.02820
  [hep-th]}}.

\bibitem{Chaudhuri:2018ymp}
S.~Chaudhuri, C.~Chowdhury, and R.~Loganayagam, ``{Spectral Representation of
  Thermal OTO Correlators},''
  \href{http://dx.doi.org/10.1007/JHEP02(2019)018}{{\em JHEP} {\bfseries 02}
  (2019) 018}, \href{http://arxiv.org/abs/1810.03118}{{\ttfamily
  arXiv:1810.03118 [hep-th]}}.

\bibitem{Chaudhuri:2018ihk}
S.~Chaudhuri and R.~Loganayagam, ``{Probing Out-of-Time-Order Correlators},''
  \href{http://dx.doi.org/10.1007/JHEP07(2019)006}{{\em JHEP} {\bfseries 07}
  (2019) 006}, \href{http://arxiv.org/abs/1807.09731}{{\ttfamily
  arXiv:1807.09731 [hep-th]}}.

\bibitem{Kubo:1957mj}
R.~Kubo, ``{Statistical mechanical theory of irreversible processes. 1. General
  theory and simple applications in magnetic and conduction problems},''
  \href{http://dx.doi.org/10.1143/JPSJ.12.570}{{\em J. Phys. Soc. Jap.}
  {\bfseries 12} (1957) 570--586}.

\bibitem{Martin:1959jp}
P.~C. Martin and J.~S. Schwinger, ``{Theory of many particle systems. 1.},''
  \href{http://dx.doi.org/10.1103/PhysRev.115.1342}{{\em Phys. Rev.} {\bfseries
  115} (1959) 1342--1373}.

\bibitem{doi:10.1080/00031305.1995.10476177}
S.~Chib and E.~Greenberg, ``Understanding the metropolis-hastings algorithm,''
  \href{http://dx.doi.org/10.1080/00031305.1995.10476177}{{\em The American
  Statistician} {\bfseries 49} no.~4, (1995) 327--335}.

\bibitem{Feynman:1963fq}
R.~P. Feynman and F.~L. Vernon, Jr., ``{The Theory of a general quantum system
  interacting with a linear dissipative system},''
  \href{http://dx.doi.org/10.1016/0003-4916(63)90068-X}{{\em Annals Phys.}
  {\bfseries 24} (1963) 118--173}.
[Annals Phys.281,547(2000)].

\end{thebibliography}\endgroup



\end{document}